\documentclass[aps,prb,twocolumn,groupedaddress,showpacs]{revtex4}

\usepackage[bookmarks]{hyperref}
\usepackage[dvips]{graphicx}
\usepackage{amsmath}
\usepackage{amstext}
\usepackage{graphics}
\usepackage{exscale}
\usepackage{epsfig}
\usepackage[T1]{fontenc}
\usepackage{bm}
\usepackage{bbm}

\usepackage{version}

\usepackage{latexsym}

\renewcommand{\epsilon}{\varepsilon}

\newcommand{\gfbraket}[1]{\langle\!\langle #1  \rangle\!\rangle}

\newcommand{\unitop}{\mathbbm{1}}

\newcommand{\integral}[3]{\!\int\limits_{#2}^{#3}\!\!{\rm d}#1\;}

\newcommand{\expval}[2]{ \langle  #1 #2\ \!\! \rangle}

\newcommand{\elcre}[2]{ c^{\dagger}_{#1,#2}}
\newcommand{\elann}[2]{ c_{#1,#2}}

\newcommand{\e}{\mathrm e}

\newcommand{\vct}[1]{\bm #1}
\newcommand{\vk}{{\bm k}}

\newcommand{\ord}[1]{{\cal O}(#1)}

\newcommand{\Hand}{H_{\mathrm{And}}}
\newcommand{\thGf}{{\cal G}}

\newcommand{\Imag}{\mathrm{Im}}
\newcommand{\Real}{\mathrm{Re}}

\newcommand{\hc}{\mathrm{h.c.}}

\includeversion{commentst1} 

\begin{document}

\title{ Dynamical mean-field theory and numerical renormalization group
   study of superconductivity in the attractive Hubbard model} 
\author{J. Bauer${}^1$, A.C. Hewson${}^2$ and N. Dupuis${}^3$}
\affiliation{${}^1$Max-Planck Institute for Solid State Research, Heisenbergstr.1,
  70569 Stuttgart, Germany} 
\affiliation{${}^2$Department of Mathematics, Imperial College, London SW7 2AZ,
  United Kingdom}
\affiliation{${}^3$ Laboratoire de Physique Th\'eorique de la Mati\`ere
  Condens\'ee, CNRS - UMR 7600, Universit\'e Pierre et Marie Curie, 4 Place
  Jussieu, 75252 Paris Cedex 05, France }
\date{\today} 
\begin{abstract}
We present a study of the attractive Hubbard model based on the dynamical mean
field theory (DMFT) combined with the numerical renormalization group (NRG).
For this study the NRG method is extended to deal with self-consistent solutions of effective
impurity models with superconducting symmetry breaking. We give details of
this extension and validate our calculations with DMFT results with
antiferromagnetic ordering. We also present results for static and integrated
quantities for different filling factors in the crossover from weak (BCS) to
strong coupling (BEC) superfluidity. We study the evolution of the
single-particle spectra throughout the crossover regime. 
Although the DMFT does not include the interaction of the fermions with the
Goldstone mode, we find strong deviations from the mean-field theory in the
intermediate and strong coupling (BEC) regimes. In particular, we show that
low-energy charge fluctuations induce a transfer of spectral weight from the
Bogoliubov quasiparticles to a higher-energy incoherent hump.
 \end{abstract}

\maketitle

\section{Introduction}

The Hubbard model of locally interacting fermions plays a fundamental role in
the theory of condensed matter physics and has become a standard model
to study correlated electronic behavior. In its repulsive version depending
on interaction strength and doping it displays magnetic instabilities such as
antiferromagnetism.  However, there is also evidence \cite{HM00,ZS00,HSFR01,AAPH06a}
that there is a parameter range where it possesses a strong 
instability in the pairing channel to d-wave superconductivity, which makes it a
good candidate to describe many important aspects of the high temperature superconductors.  
Its attractive counterpart, the model with an onsite pairing term, has
a simpler phase diagram, as the ground state is an s-wave superconductor. At 
half filling a degenerate charge ordered state can also occur.
For electrons in a solid this model may seem inappropriate  at first sight, but one can think of the 
local attraction between the electrons as mediated by a boson, for instance, a phonon or
exciton, where any form of retardation is neglected.\cite{MRR90} Indeed, the
\citet*{BCS57} (BCS) theory for superconductivity 
uses a similar model with instantaneous local attraction albeit with an
energy (Debye) cutoff. In ultracold atom experiments 
\cite{BDZ08} the interactions between the fermionic atoms in an optical trap 
can be tuned by a Feshbach resonance. For a broad resonance there exists a
regime where the effective interaction is well described by a local 
attraction. Superfluidity has been observed in such systems
\cite{GRJ03,ZSSRKK04,ZSSSK05,BDZ08}, also in the case where the fermions are
confined to an optical lattice \cite{CMLSSSXK06}. 

When tuning the interaction in models of attractive fermions, such as the
attractive Hubbard model, one has two limiting cases, that of weak coupling BCS
superfluidity and  the strong coupling Bose Einstein condensation (BEC) of preformed pairs. The
theoretical understanding which has been developed  over the years is that the
properties, such as the order parameter $\Delta_{\rm sc}$ and the transition
temperature $T_c$ to the superfluid state, are connected by a smooth 
crossover, and approximate interpolation schemes between these limits have
been devised \cite{Eag69,Leg80,NS85,Ran95}. Apart from its recent experimental
realization for ultracold atoms in an optical trap
\cite{GRJ03,ZSSRKK04,ZSSSK05,BDZ08}, there is experimental evidence that this
BCS-BEC crossover has also relevance for strong coupling and high temperature
superconductors. It has been claimed that these superconductors display
properties in certain parts of the phase diagram, such as the pseudo-gap,
that can be  understood in terms of pairs, preformed  above the transition
temperature $T_c$, in contrast to the BCS picture, where the pairs no longer
exist above $T_c$.\cite{MRR90,TCC05,CLS06}

Many aspects of the attractive Hubbard model have already been
investigated\cite{MRR90,Dup04}. However, the dynamic response functions have
received fairly little
theoretical attention, and it is the predictions for these quantities through
the crossover that will be the focus of the  present paper. One particular
question concerns the fermionic excitations in the one particle spectral
functions. These are dominated by sharp Bogoliubov excitations in the weak coupling 
limit. However, at strong coupling, when the fermions are bound to pairs, we
expect a decrease of the spectral weight carried by the Bogoliubov quasiparticles.
In order to investigate in detail what happens throughout the crossover
a suitable approach to calculate dynamic quantities is required. In situations
where the momentum dependence of the self-energy is not so important, such as
in the Mott transition, the dynamical mean field theory (DMFT) has proven
to be useful as local interactions can be treated very accurately. A variety
of methods such as perturbation theory, quantum Monte Carlo, as well as exact
diagonalization (ED) and numerical renormalization group (NRG) are commonly
used to solve the effective impurity model. 
Amongst these methods the NRG is one of the more suitable ones to calculate low
temperature spectral functions. Since it was 
originally proposed by Wilson \cite{Wil75}, it has been developed constantly
over the years.\cite{BCP08} The way of calculating spectral functions has been
given a solid basis by the recent approach\cite{PPA06,WD07} based on complete basis set proposed
by Anders and Schiller \cite{AS05}. So far the NRG has, however, not been applied 
to self-consistent DMFT calculations with superconducting symmetry breaking. 
Here we will show in detail how the method can be extended to this
situation and present results for the spectral functions. Some of the results
have already been published in Ref. \onlinecite{BH09}. 
DMFT studies for the attractive Hubbard model based on other 'impurity
solvers'  have been carried out in the normal phase\cite{KMS01,CCG02},
and in the broken symmetry phase\cite{GKR05,TBCC05,TCC05}. There is also a
recent study in two dimensions with cellular DMFT \cite{KGT06}.

Our paper is organized as follows. The model and DMFT-NRG approach are
described in section II. For this calculation the DMFT-NRG approach has to be
generalized to deal with the case of a superconducting bath. This
generalization is described in detail in section III. 
There is a mapping from the negative $U$ model to the positive one when the
lattice is bipartite. In the half filled case this mapping can be used to
check the results for superconductivity with earlier DMFT-NRG calculations
with antiferromagnetic order. The mapping and comparison of the results is
given in section IV.  
In section V we compare our results for static and integrated quantities, such
as the anomalous expectation value or superfluid density,
with results based on other approximations. Finally in section VI we
present results for dynamic response functions. We focus on the
features in the one-electron spectral density. Dynamic
susceptibilities calculated with the method described here have been reported
in Ref. \onlinecite{BH09}.

\section{Model and DMFT-NRG setup}

The subject of this paper is a study of the attractive Hubbard model, which in the
grand canonical formalism reads 
\begin{equation}
H=\sum_{i,j,\sigma}(t_{ij}\elcre {i}{\sigma}\elann
{j}{\sigma}+\hc)-\mu\sum_{i\sigma}n_{i\sigma}-U\sum_in_{i,\uparrow}n_{i,\downarrow},
\label{attHub}
\end{equation}
with the chemical potential $\mu$, the interaction strength $U>0$ and  the
hopping parameters $t_{ij}$. $\elcre {i}{\sigma}$ creates a fermion at site
$i$ with spin $\sigma$, and $n_{i,\sigma}=\elcre {i}{\sigma}\elann
{i}{\sigma}$. The present calculations are confined to zero temperature,
however, an extension to finite temperature is possible.
To study superconducting order we can include an explicit superconducting
symmetry breaking term,
\begin{equation}
  \label{sbterm}
H_{\rm sc}=
\Delta^0_{\rm  sc}\sum_{\vk}[\elcre{\vk}{\uparrow} \elcre{-\vk}{\downarrow} +\hc],  
\end{equation}
with an ``external field'' $\Delta_{\rm  sc}^{0}$.
In the superconducting case in Nambu space the Green's function matrix is
given by 
\begin{equation}
  \underline {G}_{\vk}(\omega)=
\left(\begin{array}{c c}
\gfbraket{\elann{\vk}{\uparrow};\elcre{\vk}{\uparrow}}_{\omega} & 
\gfbraket{\elann{\vk}{\uparrow};\elann{-\vk}{\downarrow}}_{\omega} \\
\gfbraket{\elcre{-\vk}{\downarrow};\elcre{\vk}{\uparrow}}_{\omega} &
\gfbraket{\elcre{-\vk}{\downarrow};\elann{-\vk}{\downarrow}}_{\omega}
\end{array}\right),
\end{equation}
where we use the notation for zero temperature retarded Green's functions for
two operators $A,B$,
$\gfbraket{A;B}_{\omega}:=-i\integral{t}{}{}\theta(t)\e^{i\omega
  t}\expval{[A(t),B]}{}$ with the expectation value in the ground state
$\expval{\ldots}{}$. Upon including (\ref{sbterm}) the non-interacting Green's
function matrix $\underline G_{\vk}^0(\omega)$ has the form,    
\begin{equation}
  \underline G_{\vk}^0(\omega)^{-1}=
\left(\begin{array}{cc}
\omega-\xi_{\vk}  & \Delta^0_{\rm sc}  \\
\Delta^0_{\rm sc} & \omega+\xi_{\vk}
\end{array}\right),
\label{frGfctsc}
\end{equation}
where $\xi_{\vk}=\epsilon_{\vk}-\mu$. For the interacting system
we introduce the matrix self-energy $\underline\Sigma_{\vk}(\omega)$ such
that the inverse of the full Green's function matrix $\underline
G_{\vk}(\omega)$ is given by the Dyson equation  
\begin{equation}
  \underline G_{\vk}(\omega)^{-1}=
  \underline G_{\vk}^0(\omega)^{-1}-\underline\Sigma_{\vk}(\omega).
\label{scdyson}
\end{equation}
We employ the dynamical mean field theory to analyze the model (\ref{attHub}).
As effective impurity model we consider the attractive Anderson impurity model
in a superconducting medium, 
\begin{eqnarray}
\Hand^{\rm sc}&=&H_{\rm imp}+
\sum_{\vk,{\sigma}}\epsilon_{\vk}\elcre{\vk}{\sigma}\elann{\vk}{\sigma} 
+ \sum_{\vk,{\sigma}}V_{\vk}(\elcre{\vk}{\sigma}{d}_{\sigma} + \hc)
\nonumber \\
&&
-\sum_{\vk}\Delta_{\vk}[\elcre{\vk}{\uparrow} 
\elcre{-\vk}{\downarrow}+\elann{-\vk}{\downarrow}\elann{\vk}{\uparrow}].
\label{scAIMn}
\end{eqnarray}
where $H_{\rm imp}=\sum_{\sigma}
\epsilon_dn_{\sigma}-Un_{\uparrow}n_{\downarrow}$ with
$n_{\sigma}=d^{\dagger}_{\sigma}d_{\sigma}$ and $d_{\sigma}$ is the
fermionic operator on the impurity site. $\epsilon_{\vk}$, $V_{\vk}$ and
$\Delta_{\vk}$ are parameters of the medium . 
For the model (\ref{scAIMn}) the non-interacting  Green's function matrix has the form, 
\begin{equation}
 \underline{G}_0(\omega)^{-1}=
\omega\unitop_{2}-\epsilon_{d}\tau_3-\underline K(\omega).
\label{scG0}
\end{equation}
$\underline K(\omega)$ is the generalized matrix hybridization for the medium,
with diagonal part
\begin{equation}
  K_{11}(\omega)=\sum_{\vk}V_{\vk}^2
\frac{\omega+\epsilon_{\vk}}{\omega^2-(\epsilon_{\vk}^2+\Delta_{\vk}^2)} 
\label{k11sc}
\end{equation}
and offdiagonal part,
\begin{equation}
  K_{21}(\omega)=\sum_{\vk}V_{\vk}^2
\frac{\Delta_{\vk}}{\omega^2-(\epsilon_{\vk}^2+\Delta_{\vk}^2)} . 
\label{k21sc}
\end{equation}
For a self-consistent numerical renormalization group (NRG) calculation of an
effective impurity problem one has to (i) 
calculate the effective impurity model parameters $V_{\vk}$, $\epsilon_{\vk}$
and $\Delta_{\vk}$ in (\ref{scAIMn}) from a given input function
$\underline K(\omega)$ and (ii) map
(\ref{scAIMn}) to the so-called linear chain Hamiltonian, to which the iterative
diagonalization of the NRG can be applied. Due to the symmetry breaking the
standard formulation \cite{BCP08} needs to be extended. The details of how this can be
achieved are described in the next section. 

In the case with superconducting symmetry breaking, the effective Weiss field
is a  $2\times2$ matrix $\underline\thGf_{0}^{-1}(t)$.
The DMFT self-consistency equation in this case reads\cite{GKKR96}  
\begin{equation}
  \underline\thGf_{0}^{-1}(\omega)=\underline G(\omega)^{-1}
  +\underline\Sigma(\omega), 
\label{scselfcon}
\end{equation}
with $\vk$-independent self-energy \cite{MV89}.
Hence, we use the NRG to solve the effective impurity problem for a given medium
$\underline K(\omega)$ and calculate $\underline\Sigma(\omega)$ as detailed
in the appendix A.3. From this we can obtain the diagonal local lattice Green's
function, which for the superconducting case  takes the form 
\begin{equation}
G(\omega)
=\int d\epsilon\frac{\rho_0(\epsilon)(\zeta_{2}(\omega)+\epsilon)}
{[\zeta_{1}(\omega)-\epsilon][\zeta_{2}(\omega)+\epsilon]-\Sigma_{21}(\omega)\Sigma_{12}(\omega)},
\label{scgfct}
\end{equation}
where $\rho_0(\epsilon)$ is the density of states of the non-interacting
fermions and $\zeta_{1}(\omega)=\omega+\mu-\Sigma_{11}(\omega)$ and
$\zeta_{2}(\omega)=\omega-\mu-\Sigma_{22}(\omega)$. 
The offdiagonal part is given by
\begin{equation}
G^{\mathrm{off}}(\omega)
=\int d\epsilon\frac{\rho_0(\epsilon)\Sigma_{21}(\omega)}
{[\zeta_{1}(\omega)-\epsilon][\zeta_{2}(\omega)+\epsilon]-\Sigma_{21}(\omega)\Sigma_{12}(\omega)}.
\label{scgfctff}
\end{equation}
We denote $G_{11}=G$, $G_{21}=G^{\mathrm{off}}$ and
$G_{21}(\omega)=G_{12}(-\omega)^*$, $G_{22}(\omega)=-G_{11}(-\omega)^*$. These 
Green's functions can be collected into the matrix $\underline G$.  
Having calculated the local Green's function $\underline G$ the
self-consistency equation (\ref{scselfcon}) determines the new effective Weiss
field  $\underline\thGf_{0}^{-1}(\omega)$. We take the impurity model in the
form (\ref{scAIMn}), and identify
$\underline{G}_0(\omega)=\underline\thGf_{0}(\omega)$. Then  from equation   
(\ref{scG0}) we obtain an equation for the effective medium matrix 
$\underline K(\omega)$. In the calculations with spontaneous superconducting
order we will always consider the limit $\Delta_{\rm  sc}^0\to 0$ in equation
(\ref{sbterm}), where a solution with superconducting symmetry 
breaking will have bath parameters $\Delta_{\vk}\neq 0$ in the effective
impurity model (\ref{scAIMn}).
In section IV we compare the results of our extended method with
the ones from a well-known antiferromagnetic case in order to gauge the quality of the 
new scheme.

\section{Extension of the NRG formalism with superconducting symmetry breaking}
In this section we give details for the
extension of the DMFT-NRG calculations with superconducting symmetry
breaking. We first outline how to extract the parameters of the impurity model
from the medium function. Then we discuss the mapping to the linear chain
Hamiltonian with details in appendix A.1. This is a generalization of the scheme for the normal case
\cite{BCP08}. In the appendix A.3, we describe the generalization of the calculation of the
self-energy via the higher order Green's functions.
\subsection{Parameters of the effective impurity model}
In the self-consistent procedure the parameters of the effective impurity
model have to be determined from the input functions of the medium $K_{11}$ and
$K_{21}$, equations (\ref{k11sc}) and (\ref{k21sc}). 
We start with the Hamiltonian in the form (\ref{scAIMn}) and choose a
discretization in the usual logarithmic way to intervals $I_n^{\alpha}$,
$I^+_n=(x_{n+1},x_n)$
$I^-_n=-(x_n,x_{n+1})$, $x_n=x_0\Lambda^{-n}$,
characterized by the parameter 
$\Lambda>1$, and $x_0$ large enough to cover nonzero spectral
weight. Following the normal discretization steps\cite{BCP08} retaining only
the lowest Fourier component yields 
\begin{eqnarray}
\Hand^{\rm sc}&=&H_{\rm imp}+ \sum_{{\sigma},n,\alpha}\xi_n^{\alpha}  a_{\alpha,n,\sigma}^{\dagger} a_{\alpha,n,\sigma}  
+ \sum_{{\sigma},\alpha,n}\gamma_n^{\alpha}
(a_{\alpha,n,\sigma}^{\dagger}{d}_{\sigma}
\nonumber \\
&& + \hc) - \sum_{\alpha,n}\delta_n^{\alpha}
 (a_{\alpha,n,\uparrow}^{\dagger}a_{\alpha,n,\downarrow}^{\dagger} + a_{\alpha,n,\downarrow}a_{\alpha,n,\uparrow}).
\label{hamxigamsc2disc}
\end{eqnarray}  
We outline a procedure to obtain the parameters
$\xi_n^{\alpha}$, $\gamma_n^{\alpha}$ and $\delta_n^{\alpha}$. For the
discretized model (\ref{hamxigamsc2disc}) we find similar equations to  
(\ref{k11sc}) and (\ref{k21sc}),
\begin{eqnarray}
    K_{11}(z)&=&\sum_{n,\alpha}\gamma_n^{\alpha}\,{}^2
\frac{z+\xi_n^{\alpha}}{z^2-E_n^{\alpha}\,{}^2} ,
\label{scmedeq11disc}\\
  K_{21}(z)&=&\sum_{n,\alpha}\gamma_n^{\alpha}\,{}^2
\frac{\delta_{n}^{\alpha}}{z^2-E_n^{\alpha}\,{}^2},
\label{scmedeq21disc}
\end{eqnarray}
with $E_n^{\alpha}=\sqrt{\xi_n^{\alpha}\,{}^2+\delta_{n}^{\alpha}\,{}^2}$. 
The imaginary parts  $\Delta(\omega):=-\Imag   K_{11}(\omega+i\eta)/\pi$ and $\Delta^{\rm off}(\omega):=-\Imag
  K_{21}(\omega+i\eta)/\pi$  can be written as sums of delta functions,
\begin{eqnarray*}
  \Delta(\omega)&=&\sum_{n,\alpha}\gamma_n^{\alpha}\,{}^2
 [u^2_{n,\alpha}\delta(\omega-E_n^{\alpha})+ v^2_{n,\alpha}\delta(\omega+E_n^{\alpha})],
 \\
  \Delta^{\rm off}(\omega)&=&
  \sum_{n,\alpha}\gamma_n^{\alpha}\,{}^2u_{n,\alpha}v_{n,\alpha}
 [\delta(\omega-E_n^{\alpha})-\delta(\omega+E_n^{\alpha})],
 \end{eqnarray*}
where
\begin{equation}
  u^2_{n,\alpha}=\frac12\Big(1+\frac{\xi_n^{\alpha}}{E_n^{\alpha}}\Big)
\qquad \mathrm{and} \qquad
  v^2_{n,\alpha}=\frac12\Big(1-\frac{\xi_n^{\alpha}}{E_n^{\alpha}}\Big),
\end{equation}
with $u^2_{n,\alpha}+v^2_{n,\alpha}=1$.  We define the spectral weights for
the delta function representation in the intervals $I_n^{\alpha}$ by
\begin{equation}
w_{n,\alpha}=\integral{\omega}{I_n^{\alpha}}{}\Delta(\omega) \qquad
\mathrm{and} \qquad \bar
w_{n,\alpha}=\integral{\omega}{I_n^{\alpha}}{}\Delta^{\rm off}(\omega).
\end{equation}
If we assume that $E_n^{\alpha}\in I_n^{\alpha}$, then the equations
give for $\alpha=+$,
\begin{eqnarray}
  w_{n,+}&=&\gamma_n^{+}\,{}^2u_{n,+}^2+\gamma_n^{-}\,{}^2u_{n,-}^2,\\
  \bar w_{n,+}&=&\gamma_n^{+}\,{}^2u_{n,+}v_{n,+}+\gamma_n^{-}\,{}^2u_{n,-}v_{n,-},
\label{1stweighteq}
\end{eqnarray}
and similarly for $\alpha=-$.
This leads to three independent equations to determine the four sets of independent
parameters $\gamma_n^{+}\,{}^2$, $\gamma_n^{-}\,{}^2$, $u_{n,+}$ and
$u_{n,-}$.  Hence, we are free to choose one of them,
e.g. $\gamma_n^{+}\,{}^2=w_{n,+}$, from which follows 
directly $\gamma_n^{-}\,{}^2=w_{n,-}$. 
We are then left with the equations
\begin{equation}
  w_{n,+}-w_{n,-}=w_{n,+}(u^2_{n,+}-v_{n,+}^2)+w_{n,-}(u_{n,-}^2-v_{n,-}^2),
\end{equation} 
and
\begin{equation}
  \bar w_{n,+}=w_{n,+} u_{n,+}v_{n,+}+w_{n,-}u_{n,-}v_{n,-}.
\end{equation}
Using the equality
\begin{equation}
  (u_{n,\alpha}^2-v_{n,\alpha}^2)^2=1-4u_{n,\alpha}^2v_{n,\alpha}^2,
\end{equation}
we can derive a quadratic equation for
$d_{uv,\alpha}=u_{n,\alpha}^2-v_{n,\alpha}^2$ with the solution
\begin{eqnarray*}
&&  d_{uv,+}=\Big[2\bar w_{n,+}^2(w_{n,+}^2-w_{n,+}w_{n,-})+w_{n,+}^4+ \\
&&    w_{n,+}w_{n,-}(w_{n,+}w_{n,-}-2w_{n,+}^2) \\ 
&& +4\bar w_{n,+}^2 w_{n,+}\sqrt{w_{n,+}w_{n,-}-\bar w_{n,+}^2}\Big]/ \\
&&\Big[w_{n,+}w_{n,-}(w_{n,+}w_{n,-}-2w_{n,+}^2)+ w_{n,+}^4+4\bar w_{n,+}^2w_{n,+}^2\Big]
\end{eqnarray*}
By definition the parameters are then obtained from
\begin{equation}
  \delta_n^{\alpha}=2 u_{n,\alpha}v_{n,\alpha} E^{\alpha}_n,\qquad 
  \xi_n^{\alpha}=(u_{n,\alpha}^2-v_{n,\alpha}^2) E^{\alpha}_n.
\end{equation}
In the symmetric case, $w_{n,+}=w_{n,-}$, this simplifies to  
\begin{equation}
  u_{n,+}^2-v_{n,+}^2=\sqrt{1-\frac{\bar w_{n,+}^2}{w_{n,+}^2}}, \qquad
  2u_{n,+}v_{n,+}=\frac{\bar w_{n,+}}{w_{n,+}}.
\end{equation}
such that
\[
  \delta_n^{+}\,{}=\frac{\bar w_{n,+}}{w_{n,+}}E_n,\;\; 
  \delta_n^{-}\,{}=\frac{-\bar w_{n,-}}{w_{n,-}}E_n, \;\;
  \xi_n^{\alpha}=\alpha\sqrt{1-\frac{\bar w_{n,+}^2}{w_{n,+}^2}}E_n.
\]

Apart from the condition that it lies in the intervals $I^{\alpha}_n$,
$E^{\alpha}_n$ has not been specified, but it is reasonable to take a value in
the middle of the 
intervals, i.e. $E_n^{\alpha}=|x_n+x_{n+1}|/2>0$. With this
choice all parameters are specified numerically and the discrete model  is
determined fully by the input functions.  It can be easily checked that this
procedure simplifies to the standard procedure \cite{BCP08} in the
case without superconducting symmetry breaking. 

It is also useful to check that in the case of a mean field
superconductor\cite{SSSS92,SSSS93,YO00,CLKB04,OTH04,BOH07,HWDB08} the usual
expressions for the impurity parameters are recovered in this scheme. 
For simplicity we assume $\Delta_{\rm sc}\ll D$ in the following. Expression
(\ref{g0impsc}) for the free impurity Green's function for this model yields
for the medium functions analytically for $|\omega|>\Delta_{\rm sc}$  
\begin{equation}
  \Delta(\omega)=\frac{\Gamma}{\pi}\frac{|\omega|}{\sqrt{\omega^2-\Delta_{\rm sc}^2}}
\end{equation}
and
\begin{equation}
  \Delta^{\rm off}(\omega)=\frac{\Gamma}{\pi}\frac{\Delta_{\rm
      sc}}{\sqrt{\omega^2-\Delta_{\rm sc}^2}}.
\end{equation}
With the described procedure one finds apart from a small correction the
standard results for $\xi_n^{\alpha}$ and $\gamma_n^{\alpha}$. In addition we obtain
\begin{equation}
  \delta_n^{\alpha}\simeq \Delta_{\rm
    sc}\Big(1+\frac{(\Lambda-1)^2}{4}+\ldots\Big) +\ord{\Delta_{\rm sc}^3},
\end{equation}
where we used an expansion both in $\Delta_{\rm sc}$ and $(\Lambda-1)$.
Hence, in the continuum limit, $\Lambda\to 1$, $\delta_n^{\alpha}=\Delta_{\rm
  sc}$ comes out correctly as the constant mean field gap parameter.

\subsection{Mapping to the linear chain}
The second important step (ii) in the self-consistent NRG procedure is to map the
discretized model (\ref{hamxigamsc2disc}) to the so called linear chain model
of the form, 
\begin{eqnarray}
  \Hand&=&H_{\rm imp} +\sum_{\sigma,n=0}^N\epsilon_{n}f^{\dagger}_{n,\sigma}
  f_{n,\sigma}  
+ \sum_{{\sigma},n=-1}^N \beta_{n}
(f^{\dagger}_{n,\sigma}f_{n+1,\sigma} \nonumber \\
&&+  \hc) -\sum_{n=0}^N\Delta_{n}(f^{\dagger}_{n,\uparrow}f^{\dagger}_{n,\downarrow}+f_{n,\downarrow}f_{n,\uparrow}),
\label{linchainhamLsc}
\end{eqnarray}
with $f_{-1,\sigma}=d_{\sigma}$ and $\beta_{-1}=\sqrt{\xi_0}$,
with 
\begin{equation}
\xi_0=\sum_n(\gamma_{n}^+\,{}^2+\gamma_{n}^-\,{}^2).
\end{equation}
As usual we define the localized state
\begin{equation}
 f_{0,\sigma}=\frac1{\sqrt{\xi_0}}\sum_n(\gamma^+_{n} a_{+,n,\sigma}+\gamma^-_{n}
a_{-,n,\sigma}).
\label{f0expr}
\end{equation}

The orthogonal transformation between the two Hamiltonians needs to be more
general than in the standard case since with superconducting
symmetry breaking we have superpositions of particles and holes in the
medium. We choose the following ansatz for the transformation
\begin{equation}
  f_{n,\uparrow}=\sum_{\alpha,m}u_{\alpha,nm}a_{\alpha,m,\uparrow}-v_{\alpha,nm}a_{\alpha,m,\downarrow}^{\dagger},
\label{trafoafsc2nu}
\end{equation}
and
\begin{equation}
  f_{n,\downarrow}^{\dagger}=\sum_{\alpha,m}v_{\alpha,nm}a_{\alpha,m,\uparrow}+u_{\alpha,nm}a_{\alpha,m,\downarrow}^{\dagger},
\end{equation}
We can now derive the recursion relations for the matrix elements
and the parameters. This is done in generalization of earlier work by Bulla et
al. \cite{BPH97} and the details are given in the appendix A.1. 
We find for the parameters of the linear chain Hamiltonian (\ref{linchainhamLsc})
\begin{equation}
  \label{epsnrecscn}
  \epsilon_n=\sum_{\alpha,m}\xi_m^{\alpha}(u_{\alpha,nm}^2-v_{\alpha,nm}^2)+2\delta_m^{\alpha}u_{\alpha,nm}v_{\alpha,nm},
\end{equation}
\begin{equation}
  \label{Deltanrecn}
  \Delta_n=\sum_{\alpha,m} \delta_n^{\alpha}(u_{\alpha,nm}^2-v_{\alpha,nm}^2) -2\xi_m^{\alpha}u_{\alpha,nm}v_{\alpha,nm} 
\end{equation}
and
\begin{eqnarray}
   \beta_{n}^2&=&
 \sum_{n',\alpha}\xi^{\alpha}_{n'}{}^2(u_{\alpha,nn'}^2+v_{\alpha,nn'}^2)+
\delta_{n'}^{\alpha}{}^2(u_{\alpha,nn'}^2+v_{\alpha,nn'}^2)  \nonumber\\
&&
-\epsilon_n^2-\beta_{n-1}^2-\Delta_n^2
\label{betanrecn}
\end{eqnarray}
The recursion relations for the transformation matrix elements read
\begin{eqnarray}
   \label{urecscn}
&&{\beta_n}u_{\alpha,n+1n'}= \\
&&(\xi^{\alpha}_{n'}-\epsilon_n)u_{\alpha,nn'}
+(\delta_{n'}^{\alpha}+\Delta_n)v_{\alpha,nn'}-\beta_{n-1}u_{\alpha,n-1n'}
\nonumber
\end{eqnarray}
and
\begin{eqnarray}
\label{vrecscn}
&&{\beta_n} v_{\alpha,n+1n'}= \\
&&(\delta_{n'}^{\alpha}-\Delta_n)u_{\alpha,nn'}- 
(\xi^{\alpha}_{n'}+\epsilon_n) v_{\alpha,nn'}-\beta_{n-1}v_{\alpha,n-1n'}.
\nonumber
\end{eqnarray}

\section{Comparison with AFM DMFT-NRG results}
There is a canonical transformation which maps the attractive Hubbard model with
arbitrary chemical potential to a half-filled repulsive model with a magnetic
field \cite{MRR90},
\begin{eqnarray}
  \elcre{i}{\downarrow}=\e^{i\vct q_0 \vct R_i}b_{i,\downarrow}, && 
 \elcre{i}{\uparrow}=b^{\dagger}_{i,\uparrow}, \nonumber \\
  \elann{i}{\downarrow}=\e^{-i\vct q_0 \vct R_i}b^{\dagger}_{i,\downarrow}, && 
 \elann{i}{\uparrow}=b_{i,\uparrow} ,
\label{trafospinisospin}
\end{eqnarray}
with $\vct q_0$ such that $\e^{i\vct q_0 \vct R_i}$ changes sign from one
sublattice to another. 
At half filling the respective states with broken symmetry,
superconductivity (SC) and antiferromagnetic (AFM) order, correspond directly to
each other. Hence, the quality of our new method for the superconducting
can be tested with well-known DMFT results from the case with antiferromagnetic
ordering \cite{ZPB02,BH07c}. 

The mapping can be applied to map the corresponding effective impurity models of the two cases
onto one another and we give the details in appendix B.
Here we use the  mapping (\ref{trafospinisospin}) to relate the dynamic response
functions from the AFM and the SC case, and we focus on the integrated
spectral functions for the two calculations. 
In the antiferromagnetic case in the DMFT study we usually use the A-B
sublattice basis
$C_{\vk,\sigma}^{\dagger}=(\elcre{A,\vk}{\sigma},\elcre{B,\vk}{\sigma})$, 
\begin{equation}
  \underline {G}^{\rm AFM}_{\vk}(\omega)=
\left(\begin{array}{c c}
\gfbraket{\elann{A,\vk}{\uparrow};\elcre{A,\vk}{\uparrow}}_{\omega} & 
\gfbraket{\elann{A,\vk}{\uparrow};\elcre{B,\vk}{\uparrow}}_{\omega} \\
\gfbraket{\elann{B,\vk}{\uparrow};\elcre{A,\vk}{\uparrow}}_{\omega} &
\gfbraket{\elann{B,\vk}{\uparrow};\elann{B,\vk}{\uparrow}}_{\omega}
\end{array}\right).
\end{equation}
where $\vk$ is in the reduced Brillouin zone as we have doubled the
Wigner-Seitz cell in position space including two lattice sites.
The transformation from the attractive to the repulsive model
(\ref{trafospinisospin}) yields
\begin{eqnarray}
&&  \elann{\vk}{\uparrow}\to \elann{A,\vk}{\uparrow}+\elann{B,\vk}{\uparrow}, \\
&&  \elann{\vk}{\downarrow}\to \elcre{A,\vk}{\uparrow}-\elcre{B,\vk}{\uparrow}.
\end{eqnarray}
Since we assume N\'eel type order the quantities of the B-lattice are related
to the A type lattice with opposite spin.
We find
\begin{eqnarray*}
\gfbraket{\elann{\vk}{\uparrow};\elcre{\vk}{\uparrow}}_{\omega} &\to&
G_{A,\vk,\uparrow,\uparrow}(\omega)+G_{A,\vk,\downarrow,\downarrow}(\omega) \nonumber \\ && +G_{A,\vk,\uparrow,\downarrow}(\omega)+G_{A,\vk,\downarrow,\uparrow}(\omega).   
\end{eqnarray*}
The local lattice Green's function for the antiferromagnetic Green's function
is obtained by $\vk$-summation over the reduced Brillouin zone
$\sum_{\vk}\to\integral{\epsilon}{}{} \rho_0(\epsilon)/2$,
\begin{equation}
  G_{A,\uparrow,\uparrow}(\omega)=\frac12\integral{\epsilon}{}{} \rho_0(\epsilon)
\frac{\zeta_{A,\downarrow}(\omega)}{\zeta_{A,\uparrow}(\omega)\zeta_{A,\downarrow}(\omega)
  -\epsilon^2},
\end{equation}
where $\zeta_{\alpha,\sigma}(\omega)=\omega+\mu_{\sigma}-\Sigma_{\alpha,\sigma}(\omega)$.
The offdiagonal elements vanish as product of a symmetric and antisymmetric function,
\begin{equation}
  G_{A,\uparrow,\downarrow}(\omega)=\frac12\integral{\epsilon}{}{} \rho_0(\epsilon)
\frac{\epsilon}{\zeta_{A,\uparrow}(\omega)\zeta_{A,\downarrow}(\omega)
  -\epsilon^2} =0.
\end{equation}
As a result, we can directly relate the diagonal local lattice Green's function $G_{11}(\omega)$
of the superconducting system to the sublattice Green's functions of the
antiferromagnetic system,
\begin{equation}
  G_{11}(\omega)=G_{A,\uparrow,\uparrow}(\omega)+ G_{A,\downarrow,\downarrow}(\omega).
\end{equation}
Similarly, one finds for the offdiagonal Green's function,
\begin{equation}
  G_{12}(\omega)=G_{A,\uparrow,\uparrow}(\omega)- G_{A,\downarrow,\downarrow}(\omega).
\end{equation}
The antiferromagnetic order parameter $\Delta_{\rm  AFM}=Um_A$,
$m_A=\frac12(n_{A,\uparrow}-n_{A,\downarrow})$, is therefore directly related
to the superconducting order parameter 
$\Delta_{\rm sc}=U\Phi_{\rm}$,
\begin{equation}
  \Phi=\expval{c_{0,\uparrow}c_{0,\downarrow}}{}=\integral{\omega}{-\infty}0 \Big(-\frac1{\pi}
\Imag G^{\mathrm{off}}(\omega)\Big).
\label{gapmfdef}
\end{equation}
The results in this section are calculated with the Gaussian density of
states $\rho_0(\epsilon)=\e^{-(\epsilon/t^*)^2}/\sqrt{\pi}t^*$ corresponding
to an infinite dimensional hypercubic lattice. We define an
effective  bandwidth $W=2D$ for this density of states  via $D$, the point at
which $\rho_0(D)= \rho_0(0)/\e^2$,  giving $D=\sqrt2 t^*$ corresponding to the
choice in reference \onlinecite{Bul99}. We take the value $W=4$.  

In the following Fig. \ref{statexpcomp} we show the comparison of the
anomalous expectation value $\Phi$ (SC case) with the sublattice magnetization
$m_A$(AFM case). 
 
\begin{figure}[!htbp]
\centering
\includegraphics[width=0.45\textwidth]{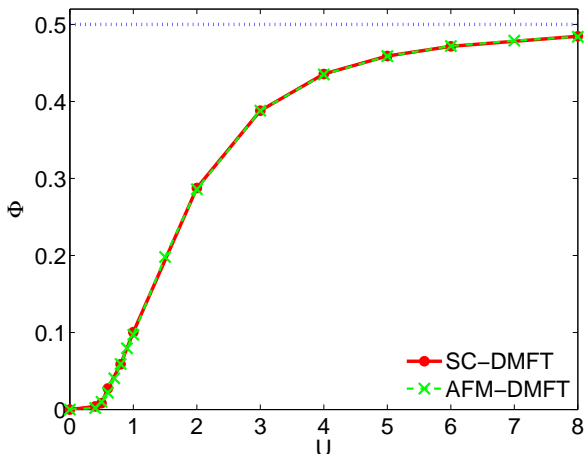}
\vspace*{-0.5cm}
\caption{(Color online) Comparison of the anomalous expectation value $\Phi$
  in the attractive model with the local magnetization $m_A$ in the AFM DMFT
  calculations for half filling.} 
\label{statexpcomp}
\end{figure}
\begin{figure}[!htbp]
\centering
\includegraphics[width=0.45\textwidth]{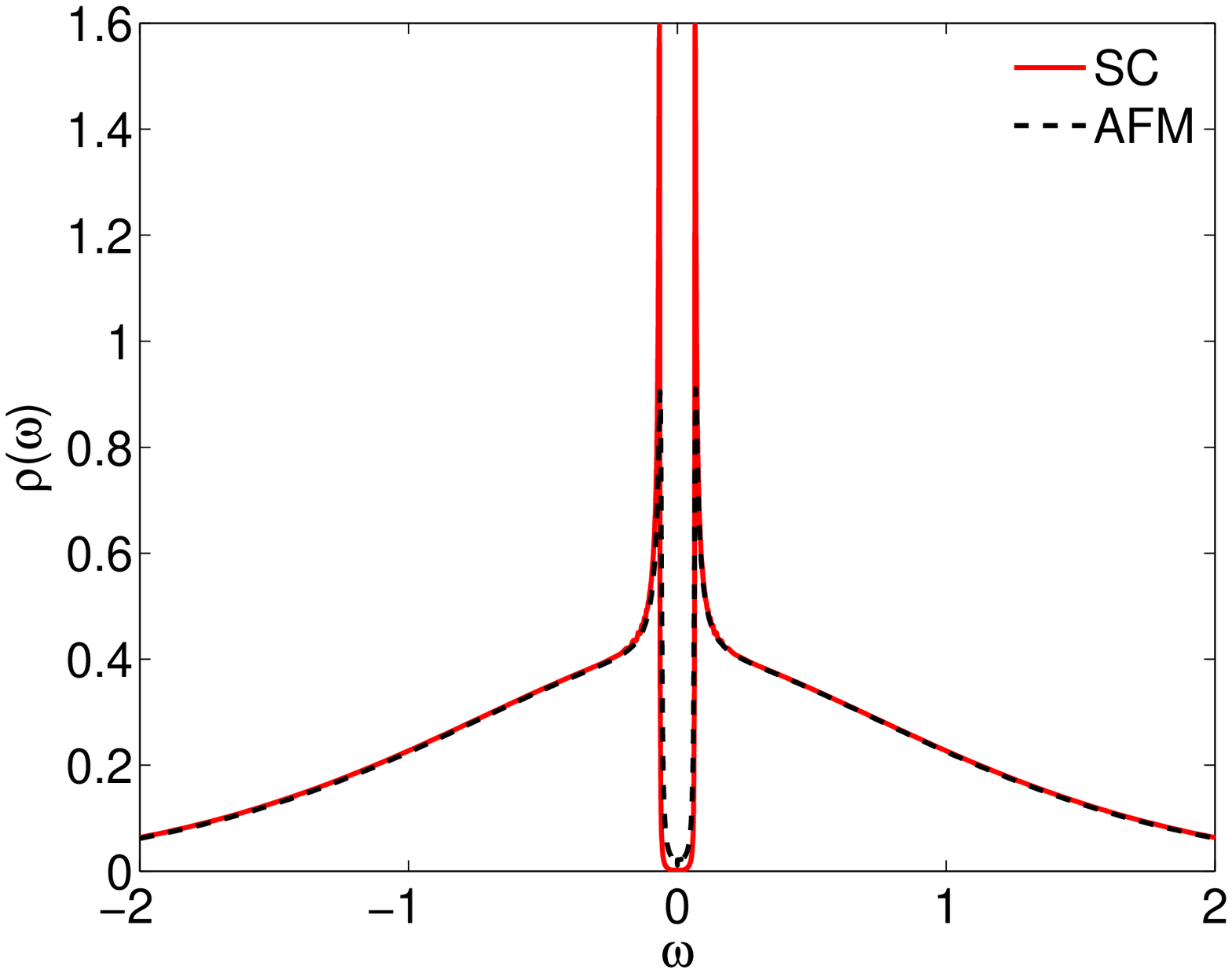}
\includegraphics[width=0.45\textwidth]{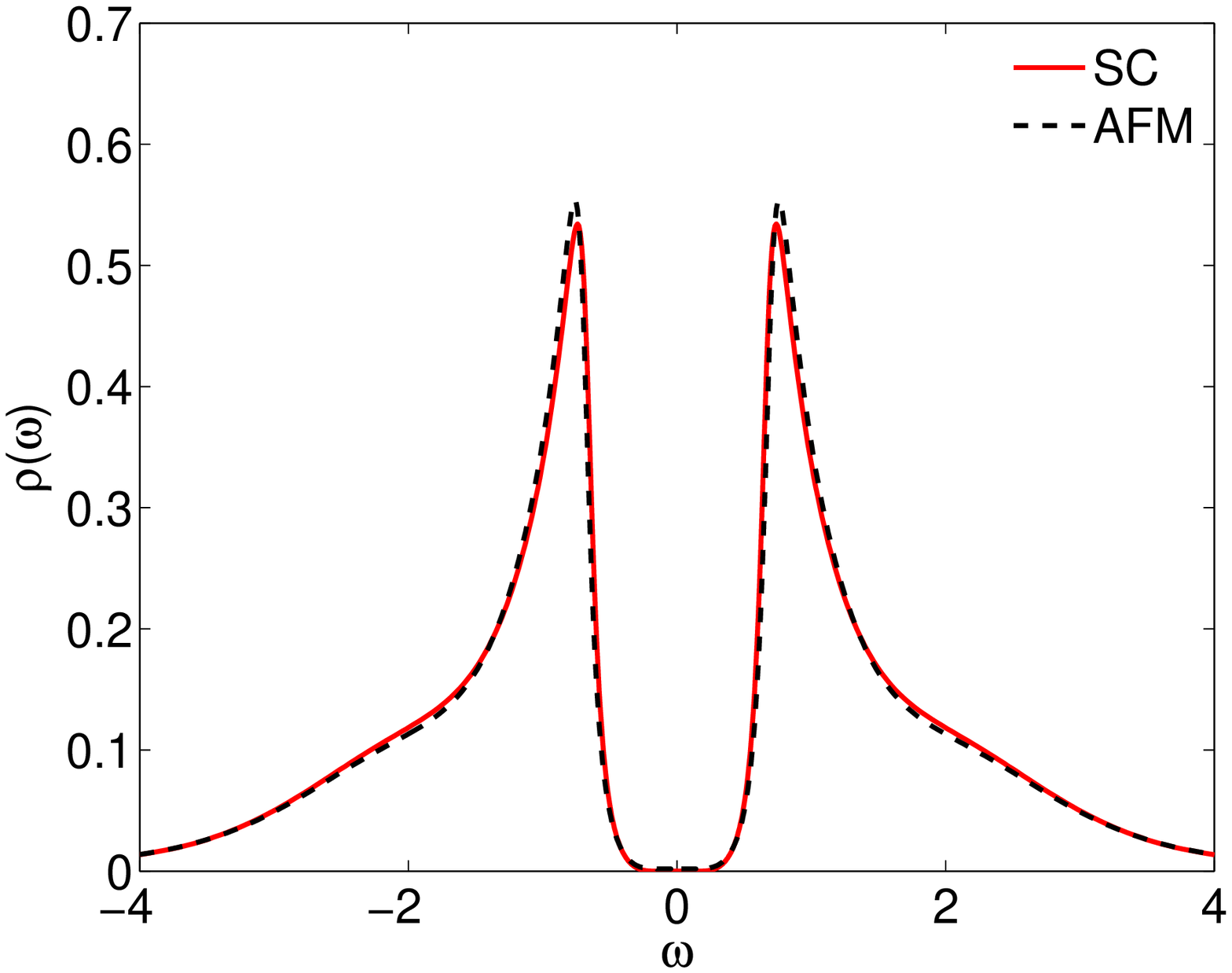}
\includegraphics[width=0.45\textwidth]{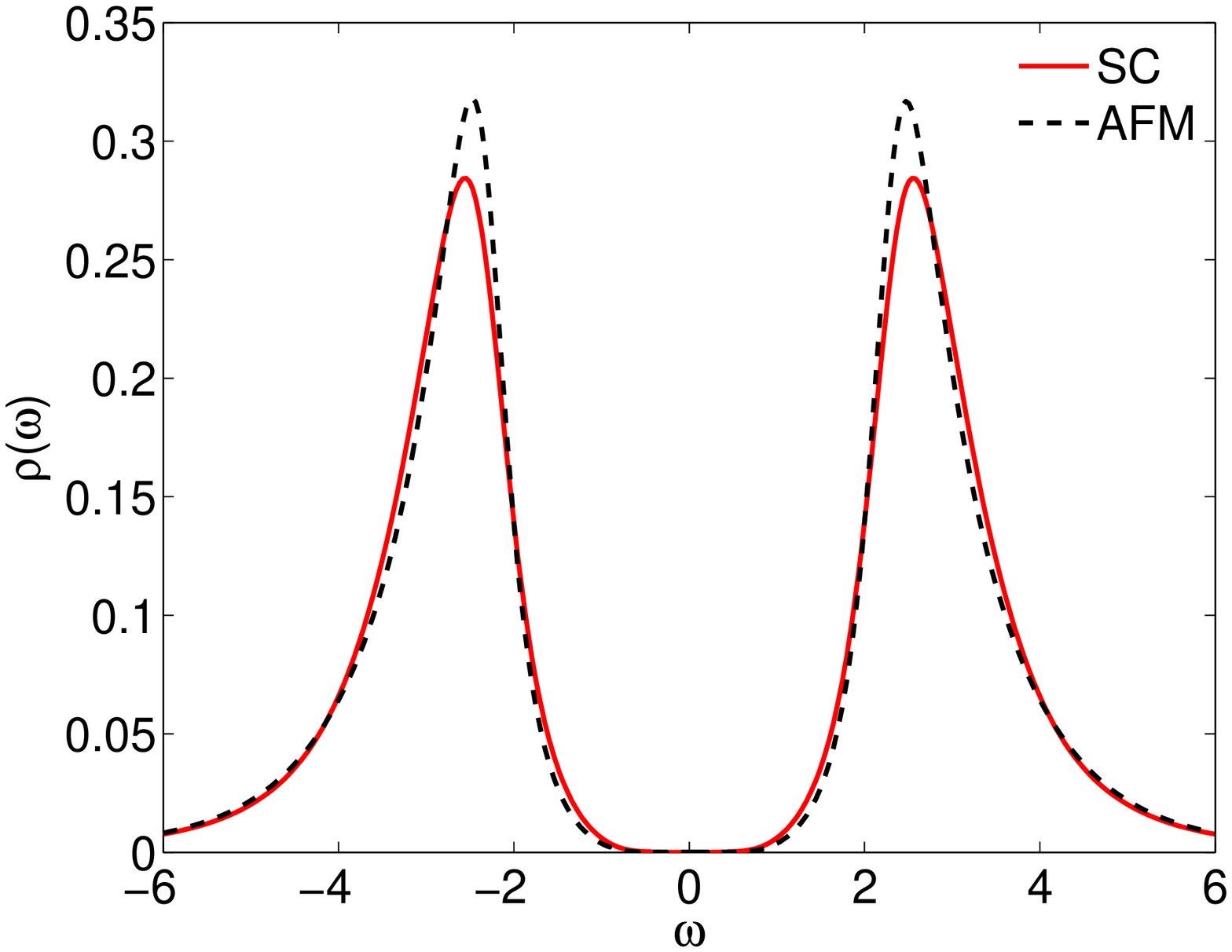}
\vspace*{-0.5cm}
\caption{(Color online) Comparison of the spectral functions of SC-DMFT and
  AFM-DMFT calculations for $U=1,3,6$ (top, middle, bottom) for half filling.}
\label{speccompafmsc}
\end{figure}

\noindent
We can see an excellent agreement of the corresponding expectation values from
the two different calculations in all coupling regimes. 
In Fig. \ref{speccompafmsc} we show the comparison for Green's functions for
$U=1,3,6$. 

\noindent
We can see that for the whole frequency range the overall agreement of these
spectral functions is good. In the weak coupling case, $U=1$, differences can
be seen in the height 
of the quasiparticle peaks, which are sharper and higher in the calculation
with superconducting order. In contrast, at strong coupling, $U=6$, the
peaks are a bit broader and not as high as in the antiferromagnetic
solution. It should be mentioned that for large $U$ DMFT-ED calculations in
the AFM state have revealed spin polaron fine structure in the peaks 
\cite{STKHCCGMAKSYSM06}. These have so far escaped NRG calculations with less
resolution at higher energy, but improved schemes might see this in the
future. Generally, the results convey the picture of a good agreement  
for static and dynamic quantities for these two different DMFT-NRG calculations.

\section{Results for static and integrated quantities}
Having tested the method at half filling we discuss results for different
filling factors in this section. We present results for static and integrated
quantities obtained with the extended DMFT-NRG method. They can be compared to
the quantities obtained with DMFT calculations with other impurity solvers,
like iterated perturbation theory (IPT)\cite{GKR05} or ED\cite{TCC05}.  
The semielliptic density of states with finite bandwidth $2D$ was used for all the
following calculations,
\begin{equation}
\rho_0(\epsilon)=\frac{2}{\pi D^2}\sqrt{D^2-\epsilon^2},
\label{dos}
\end{equation}
with $D=2t$ for the Hubbard model. $t=1$ sets
the energy scale in the following.
All the results presented here are for $T=0$. For many of the calculations
we take the model at quarter filling ($n=1/2$) as a generic case to analyze. For the NRG calculations
we use $\Lambda=1.6$ and we keep 1000 states at each step.
In the given units $U_c=2$ is the critical interaction for bound state formation in the two-body
problem for the Bethe lattice \cite{GKR05}, and can be referred to as unitarity
in analogy to the crossover terminology of the continuum system.

A starting point for an analysis of many quantities in the BCS-BEC crossover in the
attractive Hubbard model can be mean field (MF) theory.\cite{MRR90} For a given
$U$ and filling factor $n$ the chemical potential $\mu_{\rm
  MF}$ and the order parameter $\Delta_{\rm sc,MF}=U\Phi_{\rm MF}$ is
determined by the mean field equations. The fermionic excitations are given by
$E^0_{\vk}=\sqrt{(\epsilon_{\vk}-\bar\mu)^2+\Delta_{\rm sc,MF}{}^2}$ with
$\bar\mu=\mu_{\rm MF}+Un/2$.
At weak coupling the MF equations give the typical exponential behavior for $\Phi_{\rm
  MF}$, and for large $U$ one finds 
\begin{equation}
  \mu_{\rm MF}\simeq-\frac 12 U,\qquad
\Phi_{\rm  MF}\simeq\frac{\sqrt{n(2-n)}}2.
\label{mfresultmuphi}
\end{equation}
If $\bar \mu$ is larger
than the lower band energy (in 
our case $-D=-2$) then the minimal excitation energy is $\Delta_{\rm sc,MF}$
and  occurs for $\epsilon_{\vk}=\bar\mu$, which usually applies for weak
coupling. For strong coupling and $n\simeq 1$ the minimal
excitation energy is also given by $\Delta_{\rm sc,MF}$, which is of order
$U$. However, for low density, $n\to 0$, (\ref{mfresultmuphi}) yields $\bar
\mu\to - U/2$, whereas $\Phi_{\rm  MF}$ and thus $\Delta_{\rm sc,MF}$ are
small.  Once $\bar \mu$ has become smaller than the lower band
energy, the minimal excitation energy is still of order $U$ as
$E^0_{\min}=\sqrt{\bar\mu^2+\Delta_{\rm  sc,MF}^2}=U$ independent of $n$. In
the low-density strong-coupling limit the excitation gap is given by $\bar\mu$
which then corresponds to the energy of the two-fermion bound state.  

The mean field spectral densities are given  by
\begin{eqnarray}
\rho^{\rm MF}_{\vk}(\omega)&=&
{u_{\vk}^2}\delta(\omega-E^0_{\vk})+ {v_{\vk}^2}\delta(\omega+E^0_{\vk}) ,
\label{scqpgfct} \\
\rho^{\rm MF,off}_{\vk}(\omega)&=&
{u_{\vk}v_{\vk}}[\delta(\omega-E^0_{\vk})-\delta(\omega+E^0_{\vk})], 
\label{scqpgfctof}
\end{eqnarray}
where $u_{\vk}^2=(1+(\epsilon_{\vk}-\bar\mu)/{E^0_{\vk}})/2$,
$v_{\vk}^2=1-u_{\vk}^2$.  There are two bands of quasiparticle
excitations given by $\pm E_{\vk}^0$, with weights $u_{\vk}^2$ for
particle-like  and $v_{\vk}^2$ for the hole-like excitations 
with infinite lifetime. 

\subsection{Behavior of the chemical potential}
\begin{figure}[!htbp]
\centering
\includegraphics[width=0.45\textwidth]{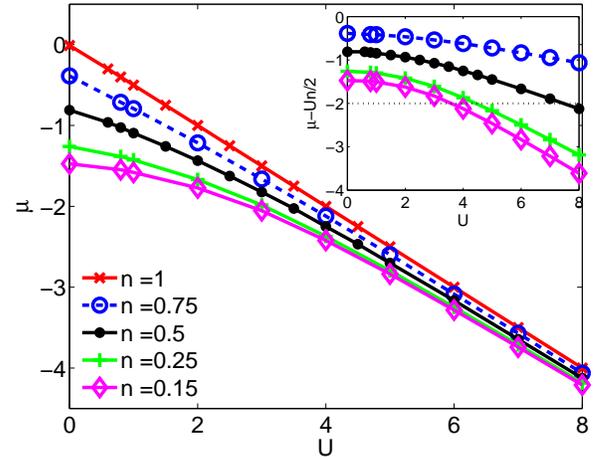}
\vspace*{-0.5cm}
\caption{(Color online) The chemical potential $\mu$ as a function of $U$ for
  different filling factors $n$. The inset shows the quantity $\mu-Un/2$.}   
\label{mu_Udep_x0.5}
\end{figure}

In Fig. \ref{mu_Udep_x0.5} we plot our DMFT results for the chemical
potential $\mu$ as a function of $U$ for different densities $n$.

\noindent
We can see that in all cases the values tend to the mean field value of $-U/2$
for large $U$. The results are in agreement with the ones reported by Garg et
al.\cite{GKR05}, and as seen there also  to the mean field values, which we
did not include in the figure.

In the inset we show the quantity $\mu-Un/2$, which corresponds to $\bar\mu$
in the mean field theory. When the density is low, e.g. $n=0.15$, it is seen
to intersect with the lower band edge $-2$ at intermediate interactions, $U\simeq
3.6$. Hence $\mu$ plays a role to determine the fermionic excitation spectrum
as discussed before. If its value does not 
change much with temperature, and $\mu-Un/2$ remains smaller than $-D$, then
no Fermi surface exists above $T_c$, and the system 
does not possess fermionic character anymore as fermions are bound to composite pairs
also above $T_c$. For large $U$, $\mu\sim -U/2$ gives the binding energy.

\subsection{Anomalous expectation value}
\begin{figure}[!htbp]
\centering
\includegraphics[width=0.45\textwidth]{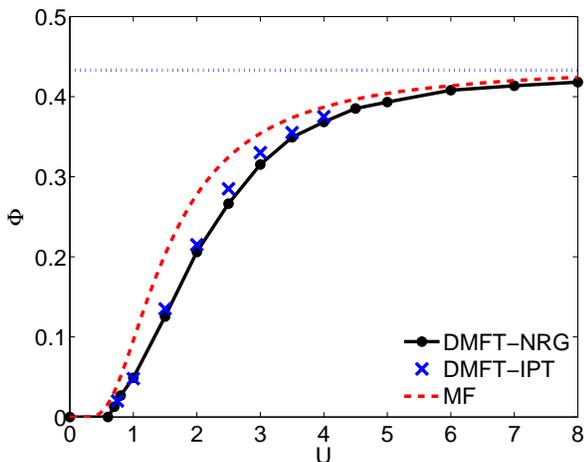}
\vspace*{-0.5cm}
\caption{(Color online) The anomalous expectation value  $\Phi$ as a function of $U$ for
  $n=0.5$. For comparison we have included the results from DMFT-IPT extracted
  from Ref. \onlinecite{GKR05} and the
  dashed line gives the result for $\Phi_{\rm MF}$.}       
\label{statexpvalsc}
\end{figure}

One of the characteristic quantities of the superconducting state is the
presence of a finite anomalous expectation value $\Phi$. The mean field equation
gives an exponential increase for $\Phi$ at weak coupling, and a quantity
which only depends on the density $n$ (\ref{mfresultmuphi}) in the strong
coupling limit. In the attractive Hubbard model the $T_c$ increases
exponentially with $U$ and then decreases at strong 
coupling with $t^2/U$ due to the kinetic term for hopping of fermionic
pairs. This is captured in the DMFT calculation, which investigates the transition
temperature as a pairing instability from the 
two particle response function.\cite{KMS01} 
We expect the anomalous expectation
value $\Phi$ in the strong coupling limit to be reduced from the mean field
value due strong phase fluctuations. This is analogous to the reduction of the
antiferromagnetic order parameter in the Heisenberg model by (transverse) spin
waves.  
The latter are however not captured within our DMFT calculations
in the state with broken symmetry, and $\Phi$ increases to a constant as in
the mean field theory, as can be seen in Fig. \ref{statexpvalsc} for quarter
filling.

\noindent
The order parameter $\Delta_{\rm   sc,DMFT}=U\Phi_{\rm DMFT}$ can,
however, be interpreted as a high energy scale for pair formation then.\cite{Ran95} 
The DMFT results for $\Phi_{\rm DMFT}$ are obtained by integration of the
offdiagonal Green's function as in equation (\ref{gapmfdef}) or the static
expectation values calculated in the NRG procedure, the results of which are
in very good agreement.   
MF and DMFT results show qualitatively a very similar overall behavior. There is a
substantial reduction of the value through the quantum 
fluctuations included in the DMFT-NRG result, which appear most pronounced in
the intermediate coupling regime, near unitarity $U_c=2$. However, also at
weak coupling there is already a correction to the mean field results. For
instance at $U=0.7$ we find $\Phi_{\rm MF}/\Phi_{\rm DMFT}\approx 2.58$. This
is comparable to the reduction found in the analysis of
Mart\'in-Rodero and Flores \cite{MF92} with second order perturbation theory.
Below $U=0.5$ the ordering scale is very small, and we do not find a well
converged DMFT solution with symmetry breaking any more. In
Fig. \ref{statexpvalsc} we have also included the results obtained by
DMFT-IPT\cite{GKR05}, which are slightly larger but otherwise in good
agreement with our DMFT-NRG results.

\subsection{Pair density}
\begin{figure}[!htbp]
\centering
\includegraphics[width=0.45\textwidth]{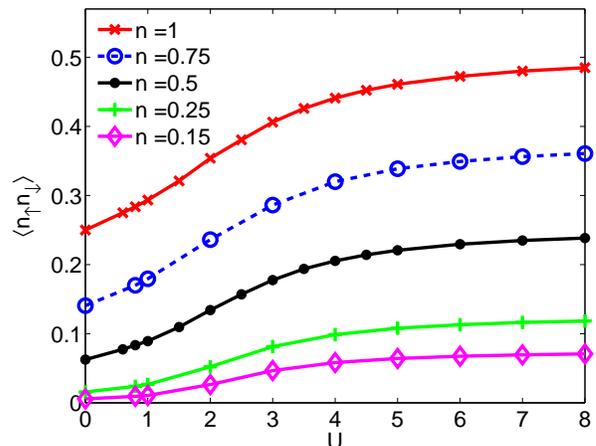}
\vspace*{-0.5cm}
\caption{(Color online) Average pair density $\expval{n_{\uparrow}n_{\downarrow}}{}$ as a
  function of $U$ for a   number of different filling factors.}
\label{docc}
\end{figure}

The ground state of the system is also characterized by the double occupancy
$\expval{n_{\uparrow}n_{\downarrow}}{}$ or average pair density. The double
occupancy multiplied by $U$ gives the expectation value of the potential
energy. At weak coupling potential energy is gained in the symmetry broken
state, whereas at strong coupling kinetic energy gain is usually responsible
for Bose Einstein condensation.  $\expval{n_{\uparrow}n_{\downarrow}}{}$ can
be calculated directly from NRG expectation values. In Fig. \ref{docc} 
it is plotted for different filling factors for a range of interactions.  

\noindent
In the non-interacting limit it is given by $(n/2)^2$, since the particles are
uncorrelated and the probabilities $n/2$ to find a particle with spin $\sigma$ 
are just multiplied. In the strong  coupling limit all particles are bound to
pairs, and the pair density is given by half the filling factor,
$\expval{n_{\uparrow}n_{\downarrow}}{}= n/2$. This continuous crossover from
the non-interacting to the strong coupling values can be seen for all densities
with the most visible change in the intermediate coupling regime occurring around $U_c=2$. 

\subsection{Momentum distribution}
\begin{figure}[!htbp]
\centering
\includegraphics[width=0.45\textwidth]{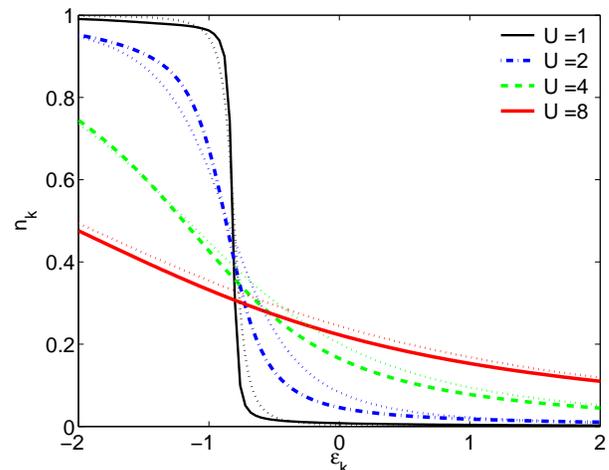}
\vspace*{-0.5cm}
\caption{(Color online) The momentum distribution calculated from the $\vk$-dependent Green's
  function and compared with the MF result $n^{\rm MF}_{\vk}=v_{\vk}^2$  (dotted
  lines) for $n=0.5$.}
\label{nkx0.5}
\end{figure}

On the mean field level the weight of the quasiparticle peaks is given
directly by the factors $u_{\vk}^2$ and $v_{\vk}^2$ as seen in equation
(\ref{scqpgfct}). These factors also describe the momentum
distribution $n^{\rm MF}_{\vk}=v_{\vk}^2$. The corresponding DMFT result for
the momentum distribution is given by the integral over the diagonal Green's
function, 
\begin{equation}
  \label{momdistr}
  n_{\vk}=\integral{\omega}{-\infty}{0}[-\Imag G_{\vk}(\omega)]/\pi.
\end{equation}
In Fig. \ref{nkx0.5} we plot the momentum distribution $n_{\vk}$ calculated
from (\ref{momdistr}) in comparison with the mean field result for $n=0.5$.   

\noindent
For small attraction ($U=1$) we can see that $n_{\vk}$ shows the typical form
known from BCS theory dropping from one to zero in a small range around
$\epsilon_{\vk}=\mu-Un/2$. Therefore, some momentum states above $\mu-Un/2$ are
occupied, but only in a small region of the size of the order parameter. When $U$ is
increased, the momentum distribution is spread over a larger range. In the BEC
limit, where the fermions  are tightly bound and therefore very localized in
position space, we expect the momentum distribution to be spread due to the uncertainty
principle. In all cases the sum rule $1/N\sum_{\vk}n_{\vk}=n/2$ is satisfied
numerically within an accuracy of about 1\%. There are visible quantitative
deviation between MF and DMFT results, but they are fairly small. Our results
comparable well to the ones presented by Garg et al. \cite{GKR05}.

In the experiments in ultracold gases where the BCS-BEC crossover is
investigated the momentum distribution can be measured quite accurately. 
This has been studied also in comparison with mean field results by Regal et
al.\cite{RGGHJ05}. Considering low densities for the lattice system, and taking
into  that an additional broadening would occur at finite temperature, a
qualitative agreement of our results with the experiment can be found.

\subsection{Superfluid stiffness}
For a system in a coherent superfluid state another characteristic quantity is the
superfluid stiffness $D_s$.  It is a measure of the energy required to 
twist the phase of the condensate, and therefore related to the degree of
phase coherence of the superconducting state. Usually, it is proportional to
the superfluid density $n_s$, which is experimentally accessible via the
penetration length. Toschi et al.\cite{TCC05} have investigated the relation
between $T_c$ and $D_s$ in the attractive Hubbard model and found that a
linear scaling relation, as in the Uemura plot, holds at intermediate and strong coupling.

$D_s$ can be calculated either from the weight of the delta-function in the
optical conductivity or from the transverse part of the current-current
correlation function\cite{TCC05} $\chi_{j_{\perp};j_{\perp}}(\vct q,\omega)$,  
\begin{equation}
  D_s=D_{\rm dia}-\chi_{j_{\perp};j_{\perp}}(\vct q\to 0,\omega=0)
\end{equation}
The diamagnetic term $D_{\rm dia}$ is essentially given by the kinetic energy,
\begin{equation}
  D_{\rm
    dia}=-\frac{2}{\beta}\sum_n\integral{\epsilon_{\vk}}{}{}\rho_0(\epsilon_{\vk})\epsilon_{\vk}  
G_{\vk}(i\omega_n),
\end{equation}
where $G_{\vk}(i\omega_n)$ is the Matsubara Green's function. 
In the infinite dimensional limit $\chi_{j_{\perp};j_{\perp}}$
reduces to the bubble of normal and anomalous propagators \cite{TCC05,PCJ93}.
From this and the relation $-\partial/\partial 
\epsilon_{\vk}[\rho_0(\epsilon_{\vk}) V(\epsilon_{\vk})] 
= \rho_0(\epsilon_{\vk})\epsilon_{\vk}$ and integration by parts 
one finds that the diamagnetic term cancels, which yields\cite{TCC05}
\begin{equation}
D_s=\frac{4}{\beta}\sum_n\integral{\epsilon_{\vk}}{}{}\rho_0(\epsilon_{\vk}) V(\epsilon_{\vk})
G_{\vk}^{\rm off}(i\omega_n)G_{\vk}^{\rm off}(i\omega_n),
\end{equation}
where $V(\epsilon_{\vk})=(4t^2-\epsilon_{\vk}^2)/3$ for the Bethe lattice.
We can use the spectral representation,
\begin{equation}
  G_{\vk}^{\rm off}(i\omega_n)=\integral{\omega'}{}{}\frac{\rho_{\vk}^{\rm
      off}(\omega')}{i\omega_n -\omega'}
\end{equation}
and the Kramers-Kronig relations for the real and imaginary parts of the Green's function
such that at zero temperature $D_s$ takes the form,
\begin{equation}
  D_s=-\frac{8}{\pi}\integral{\epsilon_{\vk}}{}{}\rho_0(\epsilon_{\vk})
 V(\epsilon_{\vk})\integral{\omega}{-\infty}0\Imag G_{\vk}^{r,\rm
   off}(\omega)\Real G_{\vk}^{r,\rm off}(\omega),   
\label{Ds1}
\end{equation}
where $ G_{\vk}^{r,\rm off}(\omega)$ is the retarded offdiagonal Green's
function (\ref{scdyson}).
We can evaluate the expression (\ref{Ds1}) using the mean field
Green's function in the form (\ref{scqpgfctof}),
which yields the somewhat simpler expression 
\begin{equation}
  D^{\rm MF}_s=4\integral{\epsilon_{\vk}}{-D}{D}\rho_0(\epsilon_{\vk})
 V(\epsilon_{\vk})\frac{u_{\vk}^2v_{\vk}^2}{E_{\vk}^0}.
\label{Ds2}
\end{equation}
This expression can be evaluated in the limit $U\to 0$, $\Delta_{\rm sc}\to
0$ as  ${u_{\vk}^2v_{\vk}^2}/{E_{\vk}^0}$ goes to a delta function then, and hence 
$D_s\to 2\rho_0(\bar\mu)V(\bar\mu)$.

\begin{figure}[!htbp]
\centering
\includegraphics[width=0.45\textwidth]{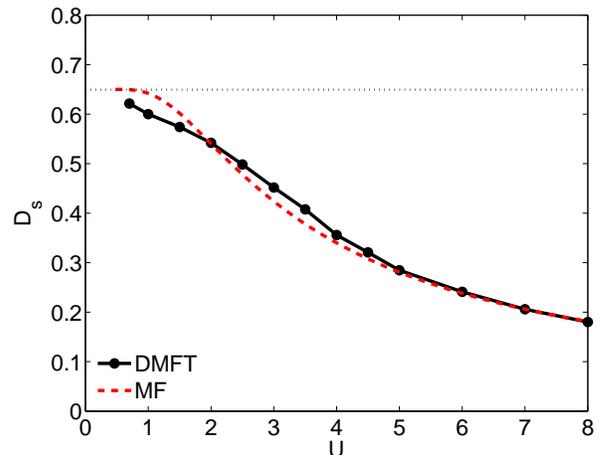}
\vspace*{-0.5cm}
\caption{(Color online) The superfluid stiffness $D_s$ as calculated from the offdiagonal
  Green's function in equation (\ref{Ds1}) for $n=0.5$. The dashed line gives
  the result for $D_s$, when evaluated as in  (\ref{Ds2}).}      
\label{DsUdep}
\end{figure}

In Fig. \ref{DsUdep} the superfluid stiffness $D_s$ calculated from equation
(\ref{Ds1}) is displayed as a function of $U$ for quarter filling.  The
dashed line shows the result as obtained from 
equation (\ref{Ds2}), where the mean field Green's functions are used to
evaluate the integrals.

\noindent
We can see that the results for $D_s$ of DMFT and MF calculation do not
deviate very much. The superfluid stiffness is maximal in the BCS limit and
decreases  to smaller values in the BEC limit. 
$D_s$ is proportional to the inverse of the effective mass of the pairs
$m_B\sim U/t^2$, and therefore expected to decrease like $1/U$. The system in this
limit consists of heavy, weakly interacting bosons, with less phase
coherence. The results shown are in agreement with the ones reported by
Toschi et al.\cite{TCC05}. 

Summarizing this section, we see that our DMFT-NRG results for chemical potential,
static and integrated properties at zero temperature are in good agreement
with earlier calculations based on different impurity solvers. In fact most of
the results 
are in good agreement with mean field theory and  quantitative
deviations due to the fluctuations included in DMFT are not very large. One
could therefore argue that the main features are already fairly well described by the
simpler static mean field treatment. In the next section we will turn to spectral
quantities. In contrast there certain features like the distinction of
coherent and incoherent excitations can only be described when we go beyond the mean
field theory. Some of these extra features found in the spectral resolution are
lost again when considering integrated quantities.

\section{Spectral functions}
In this section we present our DMFT-NRG results for the local spectral density
$\rho(\omega)$ and the $\vk$-resolved spectra, 
$\rho_{\vk}(\omega)=-\Imag G_{\vk}(\omega)/\pi$, in the different parameter
regimes.
Before discussing   these results in detail, and comparing them with those of 
 Garg et al.\cite{GKR05}, we consider the different types of approximations
used in the IPT and NRG calculations. These are relevant in  assessing the two
 sets of results to arrive at a clearer physical interpretation.

The approximation used in the IPT  is in restricting the
calculation to the second order diagram  for the self-energy. This is
 evaluated using the
Hartree-Bogoliubov corrected propagator for the effective impurity.
If this propagator has a spectral density with a gap $E_g$, then  
imaginary part of the self-energy
from this second order scattering term  can only develop in the regime  $|\omega|>3 E_g$.
This threshold energy of $3E_g$ corresponds to the minimum energy for a
 fermion above the gap $E_g$ to emit a quasiparticle-quasihole pair
 excitation. Therefore, in the IPT the single-particle spectral functions
 $\rho_{\vk}(\omega)$ have isolated
 delta-function peaks corresponding to the Bogoliubov quasiparticle excitations with minimal energy $E_g$, 
together with incoherent continuous spectrum for $|\omega|>3 E_g$. \par
 
In applying the NRG approach to the effective impurity problem,
approximations arise in using a discrete spectrum for  the
conduction electron bath. The spectral functions
are calculated as Lehmann sums over delta-function peaks, the positions
of the peaks being deduced from the discrete many-body energy levels and their weighting from the
corresponding matrix elements.
This is also the case for other methods using numerical diagonalization
such as the ED (exact diagonalization) method.  To obtain a continuous
spectral function these delta-function peaks have 
to be broadened appropriately, usually with a lognormal function with
parameter $b$ \cite{BCP08}. If the broadening is too large certain features blur, if it is
too small the spectral functions has many spikes and is difficult to
interpret. With such a broadening procedure it is difficult to resolve
sharp features such as a gap in the spectrum and hence an energy
$E_g$. However, usually an estimate of the gap can be made when the broadening
is taken into account. For all the previous results on static and integrated
quantities we have used a conventional broadening parameter
$b=0.5$, and the results for these quantities depend very
little on the broadening. In this section we  use smaller values
in order to avoid missing  features which can be lost with the larger
broadening parameter.\par

Another aspect of the NRG calculations that can lead to some numerical
uncertainty is in the way the self-energy is calculated.  In equation
(\ref{SigF}) it is shown how the self-energy can be calculated from the 
matrices of the Green's function $\underline G$ and the higher Green's
function $\underline F$. If one is
interested in the values of $\omega$ for which the imaginary part of the self-energy
vanishes then the whole expression in equation (\ref{SigF}) has to be
considered. As is well known for NRG calculations for the Anderson impurity model
\cite{BCP08} the condition $\Imag \Sigma (0)=0$  for the Friedel sum rule can be reasonably
well satisfied. However, $\Imag \Sigma$ is never exactly zero and numerical errors can often be
seen in small imaginary parts of the self-energy from this procedure. \par

\begin{figure}[!htbp]
\centering
\includegraphics[width=0.48\textwidth]{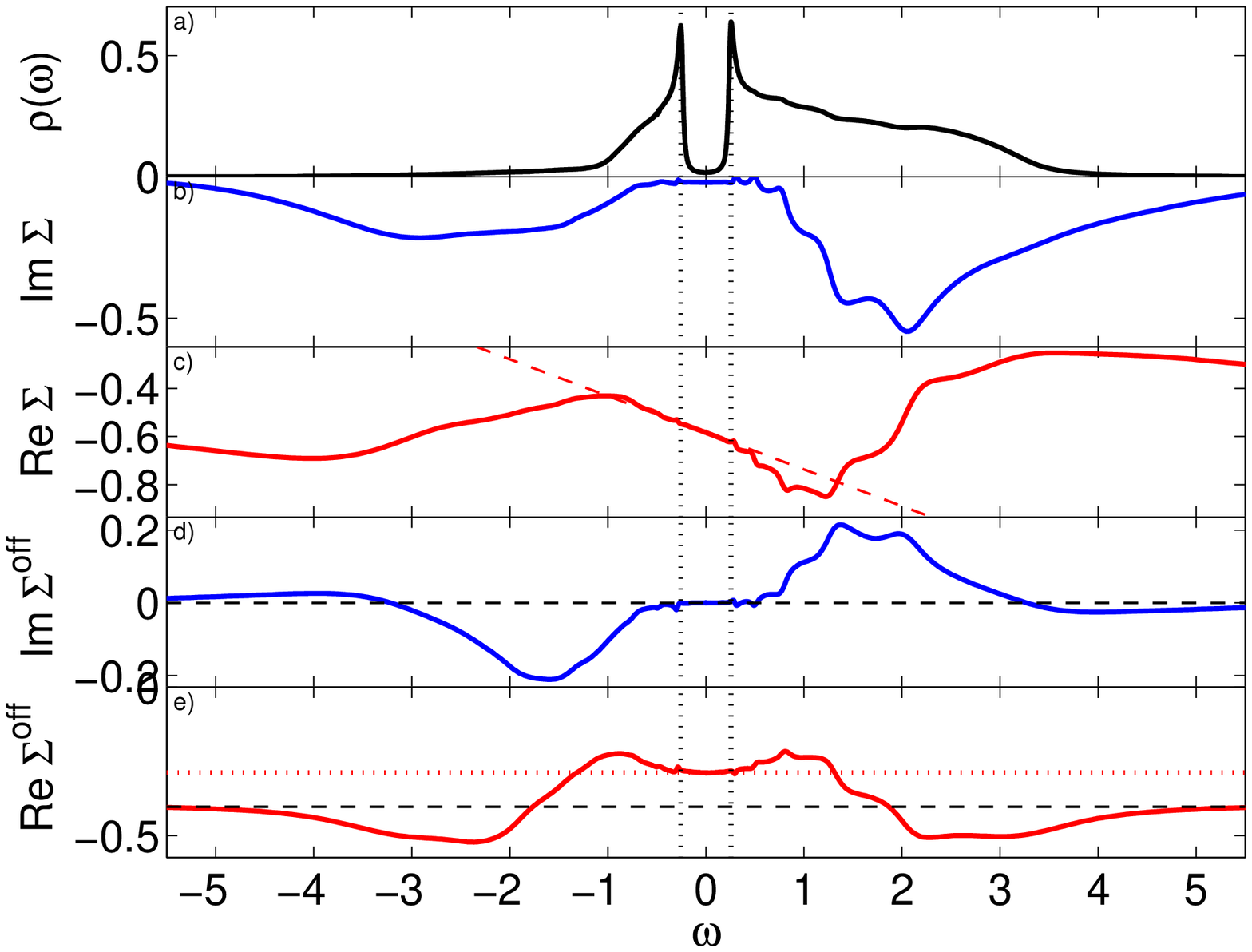}
\includegraphics[width=0.48\textwidth]{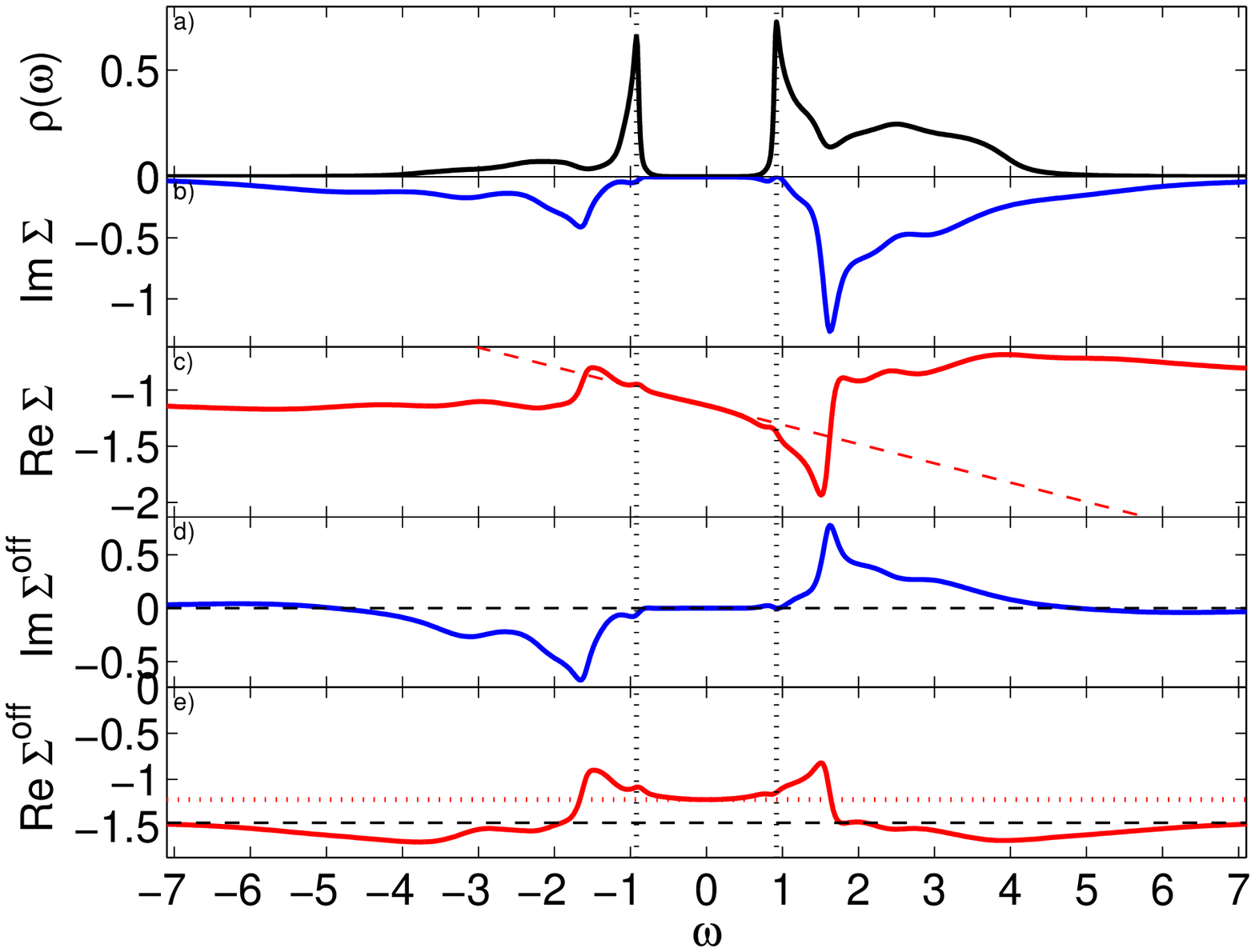}
\vspace*{-0.5cm}
\caption{(Color online) The spectral functions, imaginary and real parts of the diagonal and
  offdiagonal self-energies plotted for $U=2$ (top) and $U=4$
   (bottom) with $b=0.2$,  $n=0.5$. The dotted vertical line gives the peak
   position of the spectral function, which can be roughly identified with
   $E_g$.}     
\label{specfuncse1}
\end{figure}
\noindent
 In Fig. \ref{specfuncse1} we present the NRG results for the local spectral
 density $\rho(\omega)$ and the real and imaginary parts for the diagonal and
 offdiagonal self-energies for $U=2$ and $U=4$.
We see that  $\Imag \Sigma$ and $\Imag \Sigma^{\rm off}$  
 are approximately zero for a certain range of
$\omega$ for both for $U=2$ and $U=4$.
$\Imag \Sigma^{\rm off}$ is an antisymmetric function, which has peaks at
similar position as $\Imag \Sigma$. $\Real \Sigma^{\rm off}$ is a symmetric
function which for large $\omega$ tends to the value $\Delta_{\rm sc}=U\Phi$
of the interacting system (\ref{gapmfdef}) and for small $\omega$ can be
interpreted as a renormalized gap.   

  In the weaker coupling case $U=2$ we find a trend similar to the IPT in that $\Imag
\Sigma$  deviates appreciably  from zero only when $|\omega|>3E_g$. 
This can be seen on closer inspection of the top part of
Fig. \ref{specfuncse1}, where we estimate $E_g\approx 0.25$, and  $\Imag
\Sigma>0$ for roughly $\omega>0.75$ (the wiggles for smaller $\omega$ are
interpreted  as inaccuracies). This means that in the corresponding spectrum 
for $\rho_{\vk}(\omega)$ there are isolated quasiparticle peaks
and  a continuous incoherent part in the spectrum for $|\omega|>3E_g$.
The widths of the quasiparticles peaks, however, will not be precisely
zero as in the IPT due  to the very small imaginary parts.  In the 
 spectrum for $\rho(\omega)$ and $U=2$   shown in Fig. \ref{specfuncse1} there is a
sharp  peak due to the quasiparticle excitations in $\rho(\omega)$ just above the gap (see also Fig. 1 in
Ref. \onlinecite{BH09}). This quasiparticle band is very similar to the one
based on IPT presented in Fig. 3 in Ref. \onlinecite{GKR05}. In the IPT case
there is a square root singularity at the gap edge which does not
appear in the NRG results due to the small imaginary part in the
self-energies. 
 \par

\begin{figure}[!htbp]
\centering
\includegraphics[width=0.45\textwidth]{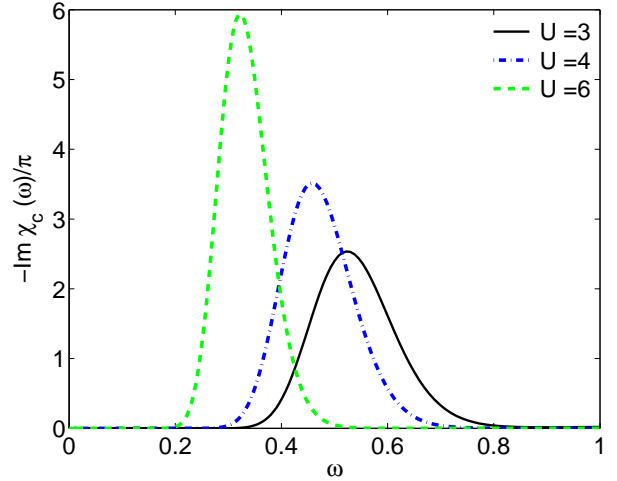}
\vspace*{-0.5cm}
\caption{(Color online) The imaginary part of the local charge susceptibility
  for $U=3,4,6$ ($b=0.2$). We find strong excitations corresponding to charge
  fluctuations there below the   respective estimates of $E_g\simeq 0.5,0.9,1.7$.}    
\label{chic_Udep_x0.5_b0.2}
\end{figure}

This picture changes in the stronger coupling case $U=4$.
Here it can be seen the imaginary parts of both the diagonal and
off-diagonal self-energies develop a pronounced peak which falls within the
region, $ E_g< |\omega| < 3E_g$. This leads to incoherent
spectral weight in $\rho_{\vk}(\omega)$ for $|\omega|<3E_g$.
This is a difference with the IPT
results where  the imaginary parts of the self-energy are always zero for
 $ E_g< |\omega| < 3E_g$ and consequently there is  no incoherent
 part of the spectrum for 
 $\rho_{\vk}(\omega)$ in the range $|\omega|<3E_g$.
 An explanation for this difference can be
seen
by examining what happens to the local dynamical charge susceptibility
$\chi_c(\omega)$ as $U$ increases.  Results for $\chi_c(\omega)$ for  $U=3,4,6$
are shown Fig. \ref{chic_Udep_x0.5_b0.2}. The excitation gap
$\Delta_c$  in this
 spectrum can be seen to decrease significantly  as $U$ increases. 
 In Ref. \onlinecite{BH09} it was found \footnote{These results are robust with 
  respect to   broadening as the excitation can be seen directly in the raw data.} 
that at strong coupling $\Delta_c$  decreases like $\sim
t^2/U$. In the weak coupling case $\Delta_c\sim 2E_g$, 
but for strong coupling $E_g$ increases with $U$, while  $\Delta_c$ decreases.
Therefore, the contribution to the self-energy arising from the scattering with
charge fluctuations can, for larger $U$,   generate a finite imaginary part
of the self-energy for $E_g<|\omega|<3E_g$.  The location of the peak in
$\Imag\Sigma$ appears consistent with such an interpretation.
The same effect cannot arise from scattering with the spin fluctuations, as these have 
a larger characteristic energy scale (of the order of $U$). For a further discussion of
the behavior of the charge and spin gap we refer to
Ref. \onlinecite{BH09}.

The development of the peak in the imaginary part of the self-energy
in the range $|\omega|<3E_g$ leads to a dip in the local spectral 
function $\rho(\omega)$ for $U=4$ as can be seen in  
Fig. \ref{specfuncse1}. There is then a peak-dip-hump structure in
$\rho(\omega)$ for $U=4$ (see also
Fig. \ref{qp_peakfit}).  This feature  has also been found in calculations for the attractive
continuum model \cite{PPS04b}. This 
behavior is not visible in the IPT calculations (cf. Fig. 3 in
Ref. \onlinecite{GKR05}). The most likely explanation is the restriction in the IPT 
to the second order diagram, which does not allow for any
renormalization of the charge fluctuations. 

\noindent

\begin{figure}[!htbp]
\centering
\includegraphics[width=0.45\textwidth]{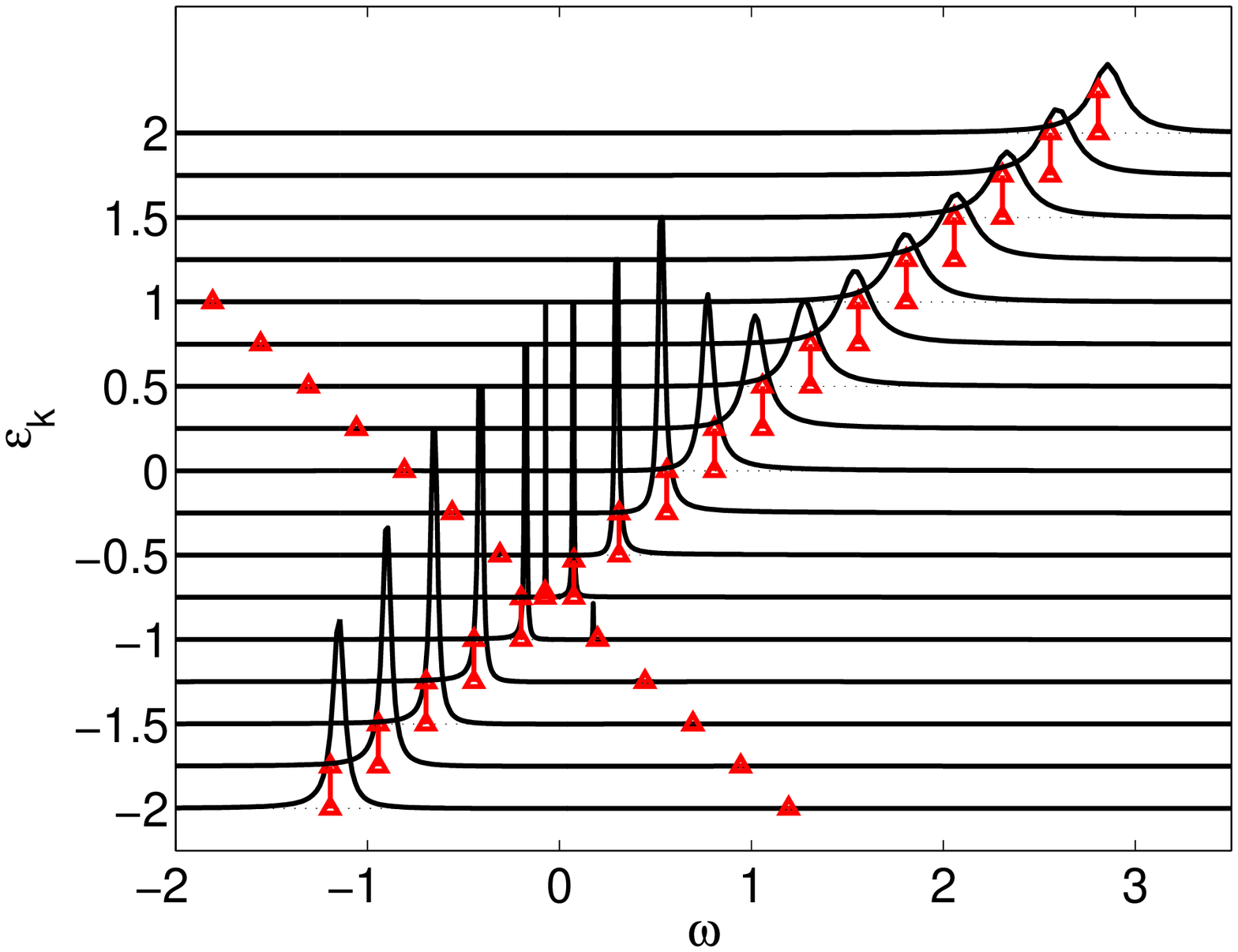}
\includegraphics[width=0.45\textwidth]{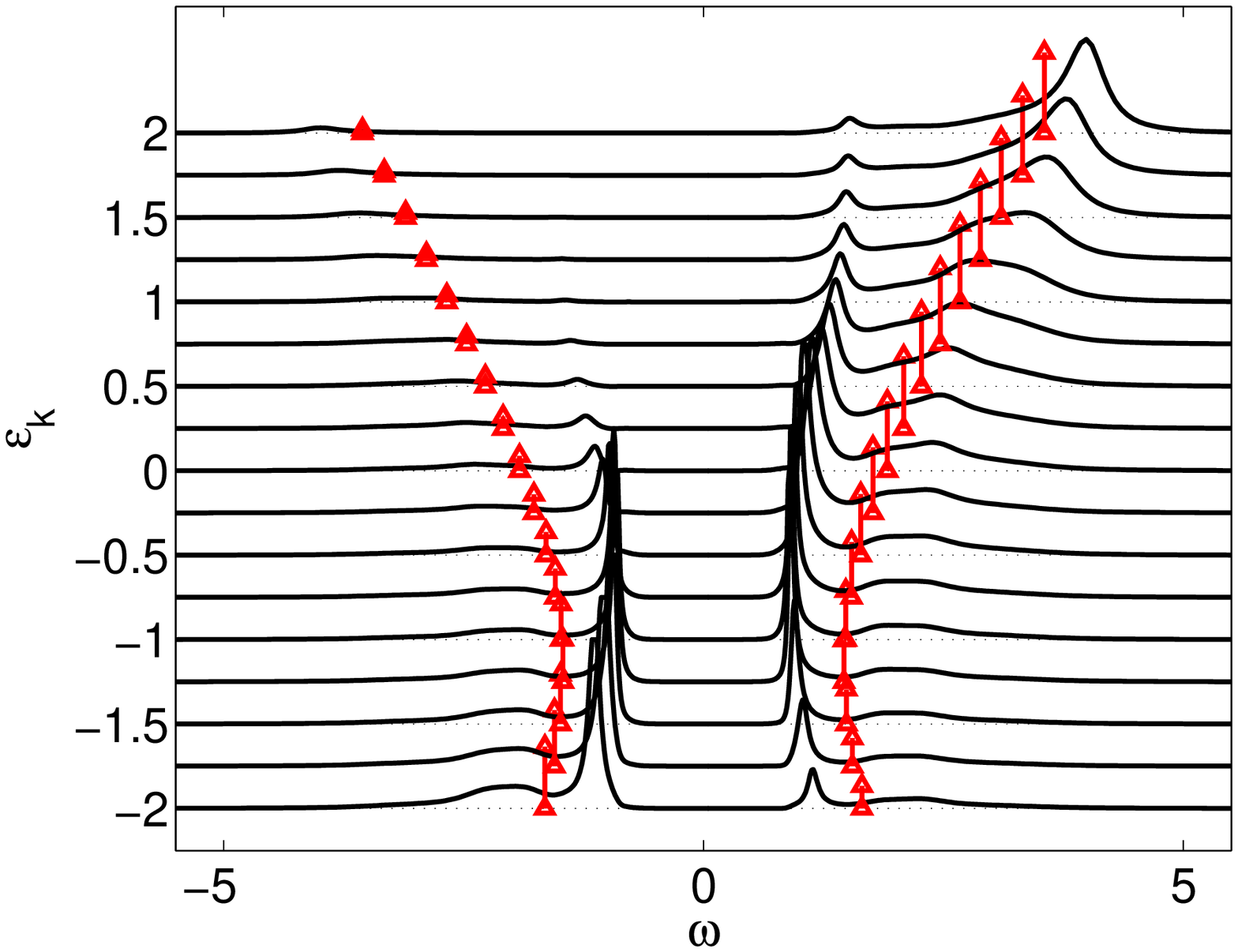}
\vspace*{-0.5cm}
\caption{(Color online) The $\epsilon_{\vk}$-resolved spectral functions $\rho_{\vk}(\omega)$ 
   for quarter filling in the BCS-limit, $U=1$ (top), and towards the BEC
   limit, $U=4$ ($b=0.2$) (bottom). The arrows show the delta-function peaks of the mean
   field solution $\rho^0_{\vk}(\omega)$, where the height of the arrow
   indicates the weight of the peak.}  
\label{kresspecscU14}
\end{figure}
\noindent
We now consider the effect of increasing $U$ on the quasiparticle
excitations which can be seen in the results for the 
spectral function $\rho_{\vk}(\omega$)  for $U=1$ and $U=4$ for
quarter filling
shown in Fig. \ref{kresspecscU14}. For both cases,  $U=1$ and $U=4$, we can
see a series of sharp quasiparticle peaks which are most narrow in 
the region $\epsilon_{\vk}\approx\bar\mu$, which is also the point where the
spectral gap is minimal. In the latter case $U=4$ in addition we find the hump with
incoherent spectral weight as discussed earlier for $\rho(\omega)$. We have
also added arrows which indicate the position of the 
quasiparticle peaks $\pm E_{\vk}^0$ in mean field theory (\ref{scqpgfct}), and
the height gives the spectral weight. We can see that they describe the
position of the quasiparticle excitation well for $U=1$. For the larger
$U$ case the structure in the spectral function is not so well captured by
the single quasiparticle peaks of the mean field theory. 
The energy of the quasiparticle excitations differ markedly from the mean
field prediction and there is a significant transfer of spectral weight to the
incoherent hump, in particular at high energy. 
The quasiparticle band (weight and band width) at low energy is, however,
captured on a qualitative level by mean field theory.
The quasiparticle peaks for $\rho_{\vk}(\omega)$ always have a finite width
since our self-energy is never strictly zero.   
One can infer the bands from the poles of the Green's function $E_{\vk}$ and
compare them with the  mean field bands $\pm E_{\vk}^0$. For the weak coupling
they are in good agreement. Towards the BEC limit 
the effective mass $m_B$ of a boson pair is of order $U$. This is reflected in
the smaller effective band width for the case $U=4$
(Fig. \ref{kresspecscU14}). The weight of the peaks in the full spectrum 
$\rho_{\vk}(\omega)$ is in  accordance with the height of the arrows 
for $\rho^0_{\vk}(\omega)$. We can see that in the BCS limit (top)
the weight in the lower band decreases rapidly to zero near
$\epsilon_{\vk}=\bar\mu$, whereas towards the BEC limit (bottom) it spreads over a
much larger region which corresponds to what has been observed for momentum
distributions in Fig. \ref{nkx0.5}.

In earlier work \cite{BH09} the quasiparticle properties were analyzed 
in an expansion around the solutions $E_{\vk}$ of the equation $\Real
G_{\vk}(\omega=E_{\vk})^{-1}=0$.  This led to the Lorentz-like quasiparticle
peak of the form 
\begin{equation}
  \rho_{\vk}(\omega)=w_+(E_{\vk})\frac{W(E_{\vk})/\pi}{(\omega-E_{\vk})^2+
    W(E_{\vk})^2},
\label{lorpeak}
\end{equation}
with width $W(E_{\vk})$ and weight $w_+(E_{\vk})$. When a standard broadening
of $b=0.5$ is used one finds a finite width of the peaks which increases in
the crossover regime leading to a strongly reduced life time of the
quasiparticles. In the more careful analysis here with smaller broadening and
taking into account possible errors in determining the self-energy we come to the
conclusion that this is not generally correct. We should not expect a finite
imaginary part of the self-energy to appear for $\omega \simeq E_g$ in a DMFT
calculation which does not include collective modes\footnote{A feature of the 
infinite dimensional model  is that it does not include a collective Goldstone
mode. A coupling of the fermions to the Goldstone mode in a more general model
can lead  to a damping of low energy  quasiparticles}.

Being aware of limitations in our numerical calculations we investigate in
more detail how the one-particle spectra near the minimal spectral gap modify
from weak to intermediate  
coupling. Here we use a general scheme in which we analyze the peaks in
the spectral function directly numerically and estimate the transfer of weight
from the quasiparticle peaks to the incoherent part of the spectrum. We take the peak position in
$\rho_{\vk}(\omega)$ for a given $\epsilon_{\vk}$ as the excitation energy
$E^{\rm ex}_{\vk}$, the full width $F_{\rm peak}$ at half maximum as the width
and the weight is  
determined by the integration over a region around $E^{\rm ex}_{\vk}$ of
$2F_{\rm peak}$. Such an analysis also applies to antisymmetric peak forms, and is
equivalent to the other one for sharp Lorentz-like peaks. Note that a
normalized Lorentz peak with width $\Delta$ (half width at half maximum)
integrated from $-2\Delta$ to $2\Delta$ yields the spectral weight
$w_{2\Delta}=2\arctan(2)/\pi\approx 0.705$.

\begin{figure}[!htbp]
\centering
\includegraphics[width=0.23\textwidth]{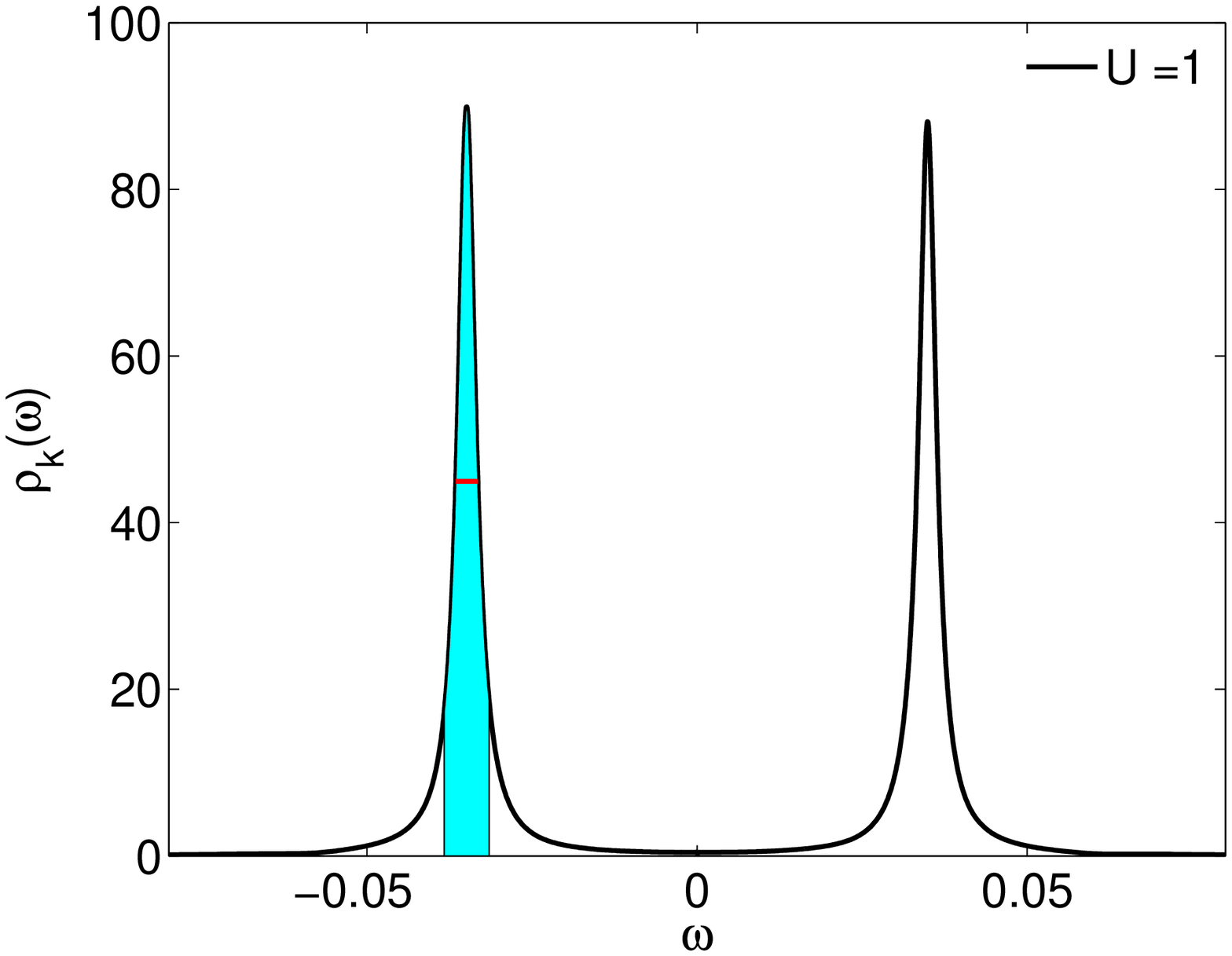}
\includegraphics[width=0.23\textwidth]{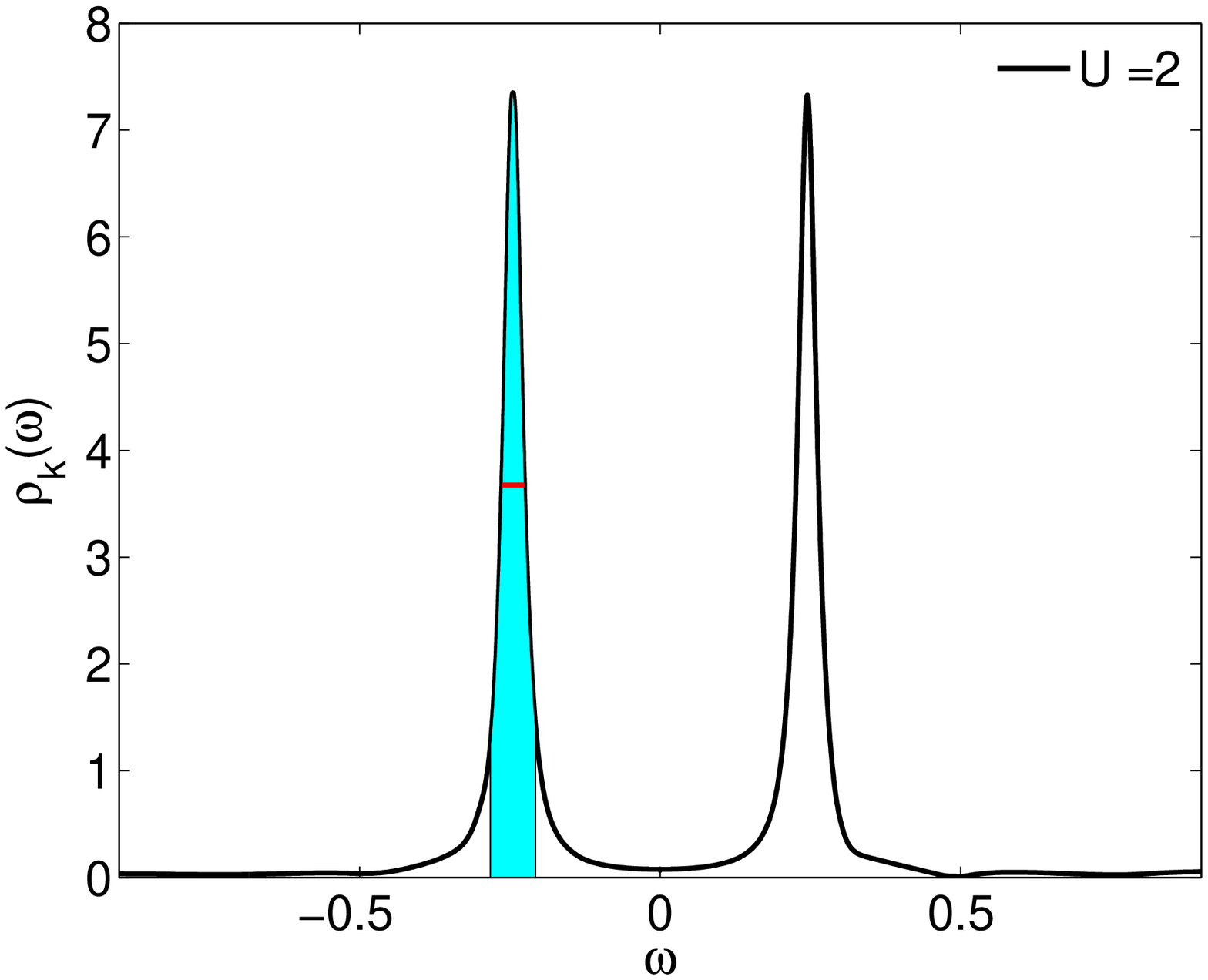}
\includegraphics[width=0.23\textwidth]{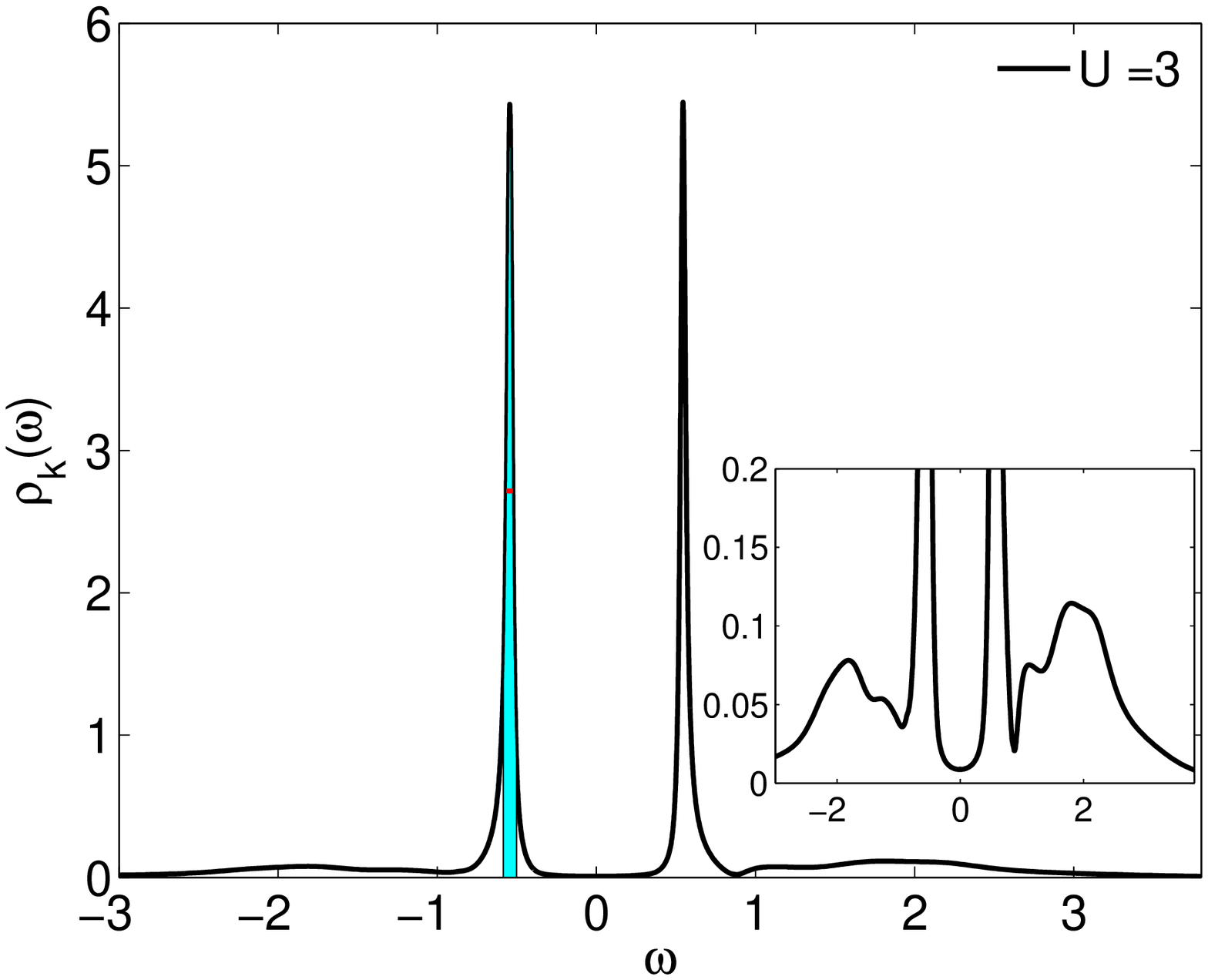}
\includegraphics[width=0.23\textwidth]{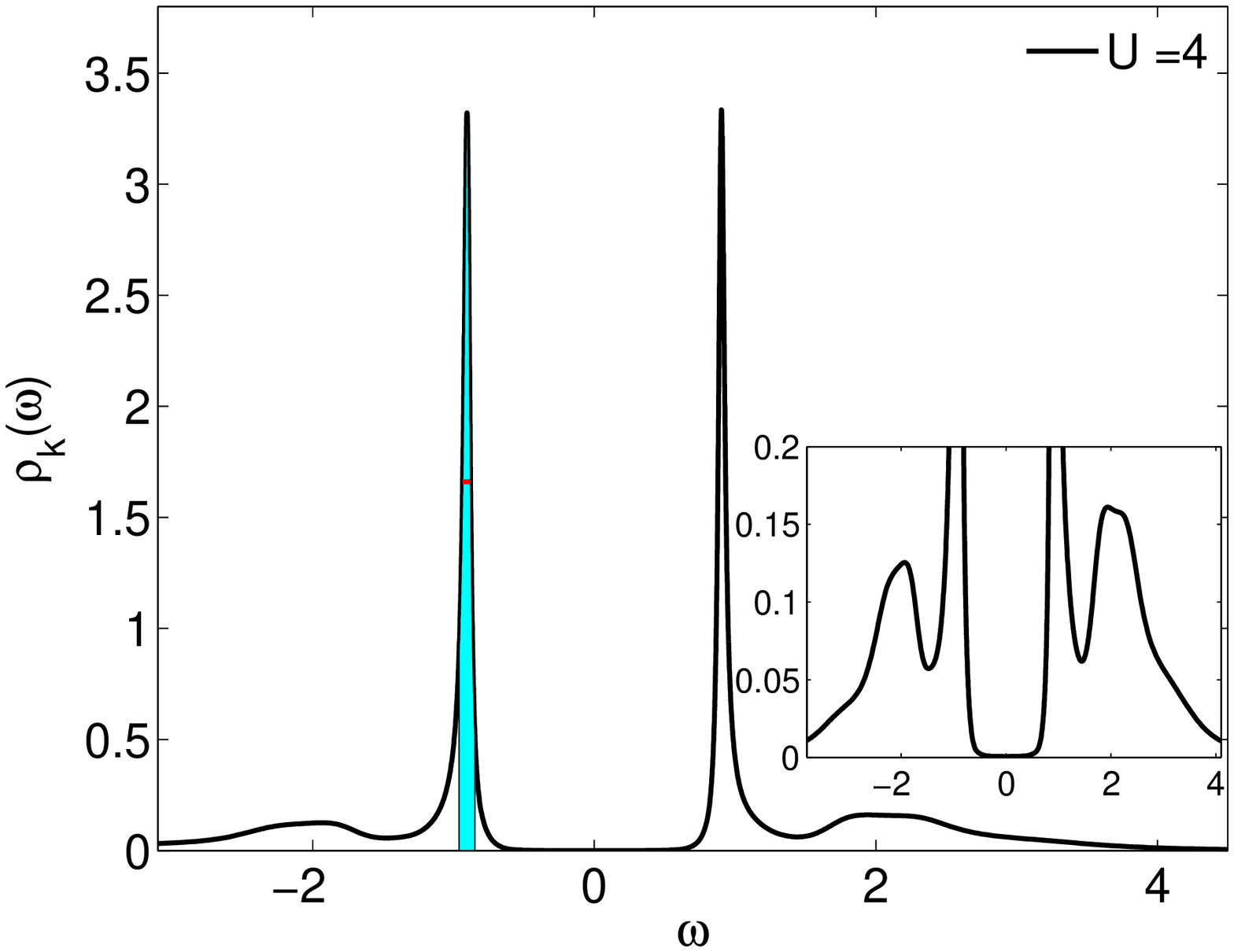}
\vspace*{-0.5cm}
\caption{(Color online) The spectral functions $\rho_{\vk}(\omega)$ for an $\epsilon_{\vk}$
  where the gap is minimal for quarter filling and $U=1-4$ ($b=0.2$). The
  integration area, which gives the weight of the peaks is shown.}  
\label{qp_peakfit}
\end{figure}
\noindent

We have done such an analysis for the $\epsilon_{\vk}$-resolved spectral
functions, where we consider an $\epsilon_{\vk}$ such that the excitation gap
is minimal. The corresponding 
spectral functions for $U=1-4$ are displayed in Fig. \ref{qp_peakfit}. 
We have included a line at half maximum for the width as well as marked the
integration area in the low energy peak.
We can see now very clearly how the spectrum changes from one coherent
quasiparticle peak at weak coupling to the peak-dip-hump structure
at intermediate coupling.  This is similar to what was found
in the calculation for an attractive continuum model \cite{PPS04b}, where a
sharp quasiparticle peak with little weight is still present at strong
coupling. In our calculation the strong coupling limit is not easy to analyze
as we always have some finite imaginary part of the self-energy leading to a
finite width of the quasiparticle peak, which is
partly spurious and tends to be larger at large coupling. At strong
coupling the excitations occur at higher energy and we have to reduce the
broadening further, which leads to a more spiky spectrum. It should be
mentioned that if the broadening is chosen larger (e.g. $b=0.5$) then there
is only one broad peak in the spectral function.

\begin{figure}[!htbp]
\centering
\includegraphics[width=0.45\textwidth]{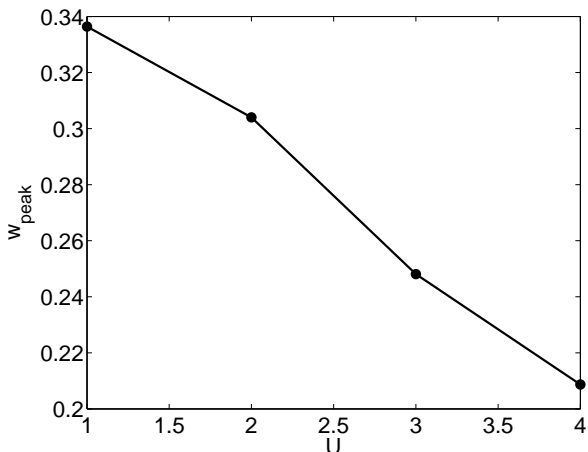}
\vspace*{-0.5cm}
\caption{The weight of the peak for spectral excitation as a
  function of $U$ for quarter filling.}    
\label{weightqpfit_Udep_x0.5}
\end{figure}

The estimate of the weight of the quasiparticle peak $w_{\rm peak}$ extracted
by integration is plotted in Fig. \ref{weightqpfit_Udep_x0.5} as a function of
$U$. For weak coupling, $U\simeq 1$, we would expect the mean field result
$v^2_{\vk}(\epsilon_{\vk}=\bar\mu)=0.5$. Due to the reduced integration range
we find $w_{\rm peak}\approx 0.34$, but division by $w_{2\Delta}$ gives a
value close to $0.5$ . 

Coming from weak coupling we find a decrease as spectral weight is
transferred to incoherent parts as seen before in Fig. \ref{qp_peakfit}. This
resembles the results of Ref. \onlinecite{PPS04b}.

As discussed before due to the uncertainty about the imaginary part of the
self-energy at low frequency the behavior of the width $F_{\rm peak}$ can not
be reliably estimated. At weak coupling we expect the prediction of a real
delta-function with $F_{\rm peak}\to 0$ to hold. Whether at strong coupling low
enough excitations can be generated to change this remains to be answered.
We stress here again that the conclusions about the behavior of
$W(\min E_{\vk})$ as given in (\ref{lorpeak}), which was reported in
Fig. 3 of Ref. \onlinecite{BH09}, are found to be incorrect as judged by
the present more careful interpretation of our DMFT-NRG calculations. 

\section{Conclusions }
In this paper we have presented an analysis of the ground state properties of
the attractive Hubbard model in the symmetry broken phase in the BCS-BEC
crossover. The emphasis has been on the evolution of spectral
functions in the crossover regime. 
Our analysis is based on an extension of the DMFT-NRG method to the case with
superconducting symmetry breaking. We have given many details of this
extension in section III and the appendix. At half filling we have related our
approach both for the effective impurity model and for the lattice quantities
to earlier DMFT-NRG calculations with antiferromagnetic symmetry breaking. A
good agreement has been found there, which validates the applicability of our
approach.  As emphasized in Ref. \onlinecite{BH09}, apart from the
attractive Hubbard model the extended method can be useful to study
superconductivity in other models, such as the Hubbard-Holstein type, and also
questions related to the microscopic description of magnetic impurities in
superconductors, which require self-consistent treatments.

We have discussed our DMFT-NRG results for static and integrated quantities,
like the anomalous expectation value, the double occupancy or superfluid stiffness.
The results for these are in good agreement with earlier calculations based on
different impurity solvers, and it has been found that most of the results are
already obtained qualitatively well on the mean field level.

The main interest of this paper has been to study the fermionic spectrum throughout
the crossover regime. The local dynamics are very well described in our
DMFT-NRG approach. We discussed how the behavior of the dynamic self-energies
changes when the interaction becomes larger. At weak coupling the spectrum is
dominated by sharp symmetric Bogoliubov quasiparticle peaks as known from mean
field theory. Contributions from particle-particle and particle-hole fluctuations
incorporated in the dynamic self-energies appear at higher energy and are
small, similar to those seen in the IPT approach. However, when the local interaction is in
the unitary regime and larger, the imaginary part of the dynamic self-energy
shows a characteristic feature which generates a peak-dip-hump structure in the
spectral function. We argued that this features is likely to be generated by
charge fluctuations as seen in the local dynamic charge susceptibility.
One finds that spectral weight is transferred into the incoherent parts (hump)
of the spectrum on increasing the coupling.   

To answer the question whether at strong coupling the fermionic quasiparticles
acquire a finite width one needs clarify over which region
the imaginary parts of the self-energies vanish. Unfortunately, our method, in
which spectral functions are obtained after broadening delta-peaks, is not
accurate enough at present to allow for definite statements. It is possible
that the peaks always remain sharp in the limit of infinite dimensions. 
Our DMFT approach does not capture spatial fluctuations and the gapless Goldstone mode. It
would also be of great interest to study how such effects give a 
modification of the discussed fermionic spectrum\cite{BD03,BD04} and possibly a suppression of
the quasiparticle peaks.   

\bigskip
\noindent{\bf Acknowledgment}\par
\noindent
We wish to thank W. Metzner, A. Oguri, P. Strack and A. Toschi for helpful discussions,
W. Koller and D. Meyer for their 
earlier contributions to the development of the NRG programs, and R. Zeyher
for critically reading the manuscript. 

\begin{appendix}
\section{NRG formalism with superconducting symmetry breaking}

\subsection{Mapping to the linear chain}
The second important step (ii) in the self-consistent NRG procedure is to map the
discretized model (\ref{hamxigamsc2disc}) to the so-called linear chain model
of the general form (\ref{linchainhamLsc}), 
\begin{eqnarray}
  \Hand&=& \sum_{\sigma,n=0}^N\epsilon_{n}f^{\dagger}_{n,\sigma}
  f_{n,\sigma}  
+ \sum_{{\sigma},n=-1}^N \beta_{n}
(f^{\dagger}_{n,\sigma}f_{n+1,\sigma} \nonumber \\
&&+  \hc) -\sum_{n=0}^N\Delta_{n}(f^{\dagger}_{n,\uparrow}f^{\dagger}_{n,\downarrow}+f_{n,\downarrow}f_{n,\uparrow}).
\label{linchainhamLscapp}
\end{eqnarray}

The orthogonal transformation has been chosen in the form (cf. equation (\ref{trafoafsc2nu})),

\begin{eqnarray}
  f_{n,\uparrow}&=&\sum_{\alpha,m}u_{\alpha,nm}a_{\alpha,m,\uparrow}-v_{\alpha,nm}a_{\alpha,m,\downarrow}^{\dagger}, 
\label{optrafo1}\\    
a_{\alpha,m,\uparrow}&=&\sum_n u_{\alpha,nm} f_{n,\uparrow}+v_{\alpha,nm}
f_{n,\downarrow}^{\dagger}, 
\label{optrafo2}\\
  f_{n,\downarrow}^{\dagger}&=&\sum_{\alpha,m}v_{\alpha,nm}a_{\alpha,m,\uparrow}+u_{\alpha,nm}a_{\alpha,m,\downarrow}^{\dagger},
  \label{optrafo3}\\ 
a_{\alpha,m,\downarrow}^{\dagger}&=&\sum_n -v_{\alpha,nm}
f_{n,\uparrow}+u_{\alpha,nm} f_{n,\downarrow}^{\dagger}.
\label{optrafo4}
\end{eqnarray}
The matrix elements of the transformation obey the relations
\begin{eqnarray*}
  \sum_nu_{\alpha,nm}u_{\alpha',nm'}+v_{\alpha,nm}v_{\alpha',nm'}&=&\delta_{m,m'}\delta_{\alpha,\alpha'}, \\
  \sum_{m,\alpha}u_{\alpha,nm}u_{\alpha,n'm}+v_{\alpha,nm}v_{\alpha,n'm}&=&\delta_{n,n'},
\end{eqnarray*}
and
\begin{eqnarray*}
  \sum_{m,\alpha}u_{\alpha,nm}v_{\alpha,n'm}-v_{\alpha,nm}u_{\alpha,n'm}&=&0,  \\
  \sum_nu_{\alpha,nm}v_{\alpha',nm'}-v_{\alpha,nm}u_{\alpha',nm'}&=&0,
\end{eqnarray*}
which ensure that both operator sets satisfy canonical anticommutation
relations. We can now derive the recursion relations for the matrix elements
and the parameters. This is done in analogy to earlier work by Bulla et
al. \cite{BPH97}. We equate the representations for the media of
(\ref{hamxigamsc2disc}) and (\ref{linchainhamLscapp}) and
substitute the operator transformation  (\ref{optrafo1})-(\ref{optrafo4}). 
One can then read off the coefficients of the $f_{n,\uparrow}$-operators
($n>0$) on both sides of the equation, which yields
\begin{eqnarray*}
&&  \sum_{n',\alpha}\xi^{\alpha}_{n'}\Big(u_{\alpha,nn'}a_{\alpha,n',\uparrow}^{\dagger}+
 v_{\alpha,nn'}a_{\alpha,n',\downarrow}\Big) \\
&& +\sum_{n',\alpha}\delta_{n'}^{\alpha}\Big(v_{\alpha,nn'}a_{\alpha,n',\uparrow}^{\dagger}-
 u_{\alpha,nn'}a_{\alpha,n',\downarrow}\Big)   = \\
&& = \epsilon_nf_{n,\uparrow}^{\dagger}+\beta_{n-1}f_{n-1,\uparrow}^{\dagger}+
  \beta_{n}f_{n+1,\uparrow}^{\dagger}-\Delta_nf_{n,\downarrow}.
\label{eq2detcoeffsc}
\end{eqnarray*}
From this we find the expression (\ref{epsnrecscn}) for $\epsilon_n$ by
taking the anticommutator with $f_{n,\uparrow}$. The anticommutator with
$f^{\dagger}_{n,\downarrow}$  gives expression (\ref{Deltanrecn}) for $\Delta_n$. 
With the representations  (\ref{optrafo1})-(\ref{optrafo4}) we can modify the
equation (\ref{eq2detcoeffsc}) to obtain
\begin{eqnarray*}
 \beta_{n}f_{n+1,\uparrow}^{\dagger}&=& \sum_{n',\alpha}\Big[(\xi^{\alpha}_{n'}-\epsilon_n)u_{\alpha,nn'}
+(\delta_{n'}^{\alpha}+\Delta_n)v_{\alpha,nn'} \\
&& -\beta_{n-1}u_{\alpha,n-1n'}\Big]a_{\alpha,n',\uparrow}^{\dagger}   \\
&+&\sum_{n',\alpha}\Big[(\Delta_n-\delta_{n'}^{\alpha})u_{\alpha,nn'}+
(\xi^{\alpha}_{n'}+\epsilon_n)
v_{\alpha,nn'}+ \\
&& \beta_{n-1}v_{\alpha,n-1n'}\Big]a_{\alpha,n',\downarrow}.
\end{eqnarray*}
By comparison with (\ref{trafoafsc2nu}) we can read off a recursion relation
for $u_{\alpha,n+1n'}$ in equation (\ref{urecscn}) and for $v_{\alpha,n+1n'}$
as in equation (\ref{vrecscn}).
The recursion relation for $\beta_n$ is obtained from the anticommutator of
with $f_{n+1,\uparrow}$ which yields
\[
  \beta_{n}^2=
  \sum_{n',\alpha}(u_{\alpha,n+1n'}^2+v_{\alpha,n+1n'}^2).
\]
With the orthonormality relations and the definitions $\epsilon_n$ and
$\Delta_n$ we can find the expression in equation (\ref{betanrecn}).


\subsection{Relevant Green's functions}
In this section we briefly outline some details for the calculations of the
relevant Green's functions and the self-energy for completeness.\cite{BOH07}
For the Green's functions it is convenient to work in Nambu space,
$\vct C^{\dagger}_{d}=(d^{\dagger}_{\uparrow},d_{\downarrow})$,
with $2\times2$ matrices.
The relevant retarded Green's functions are then 
\begin{equation}
  \underline {G}_d(\omega)=
\gfbraket{\vct C_{d};\vct C^{\dagger}_{d}}_{\omega}=
\left(\begin{array}{c c}
\gfbraket{d_{\uparrow};d^{\dagger}_{\uparrow}}_{\omega} & 
\gfbraket{d_{\uparrow};d_{\downarrow}}_{\omega} \\
\gfbraket{d^{\dagger}_{\downarrow};d^{\dagger}_{\uparrow}}_{\omega} &
\gfbraket{d^{\dagger}_{\downarrow};d_{\downarrow}}_{\omega}
\end{array}\right).
\end{equation}
In the NRG approach we calculate $G_{11}$ and $G_{21}$ directly and infer
$G_{22}(\omega)=-G_{11}(-\omega)^*$,
which follows from $G_{A,B}^{\rm ret}(\omega)=-G_{B,A}^{\rm adv}(-\omega)$ and
$G_{A,B}^{\rm ret/adv}(\omega)=-G_{A^{\dagger},B^{\dagger}}^{\rm
  ret/adv}(-\omega)^*$ for fermionic operators $A$, $B$. Similarly, we can  find
$G_{12}(\omega)=G_{21}(-\omega)^*$. In the derivation one has to be careful and include
a sign change for up down spin interchange in the corresponding operator
combination. 

In the non-interacting case we can deduce the $d$-site Green's function matrix
of the model Hamiltonian (\ref{scAIMn}) exactly. To do so we rewrite the
superconducting term of the medium $H_{\rm sc}$ by introducing the vector of
operators and the symmetric matrix 
\begin{equation}
  \label{cknambu}
\vct C_{\vk}:=
\left(\begin{array}{c}
\! \elann{\vk}{\uparrow}  \\
\! \elcre{-\vk}{\downarrow}
\end{array}\right),
\qquad
A_{\vk}:=
 \left(\begin{array}{cc}
\! \epsilon_{\vk}     & \! -\Delta_{\vk}\\
\! -\Delta_{\vk} & \! -\epsilon_{\vk}
\end{array}\right).
\end{equation}
Then $H_{\rm sc}$ can be written as
\begin{equation}
H_{\rm sc}=\sum_{\vk}\vct C_{\vk}^{\dagger}A_{\vk}\vct C_{\vk}.
\end{equation}
The matrix Green's function in the superconducting bath is then given by 
$\underline {g}_{\vk}(i\omega_n)=(i\omega_n \unitop_2-A_{\vk})^{-1}$,
\begin{equation}
 \underline {g}_{\vk}(i\omega_n)^{-1}
 =i\omega_n\unitop_{2}-\epsilon_{\vk}\tau_3
+\Delta_{\vk}\tau_1,
\label{freescbandgfct}
\end{equation}
where $\tau_i$ are Pauli matrices.
It follows that
\begin{equation}
 \underline {g}_{\vk}(i\omega_n)
 =\frac{i\omega_n\unitop_{2}+\epsilon_{\vk}\tau_3
-\Delta_{\rm sc}\tau_1}{(i\omega_n)^2-(\epsilon_{\vk}^2+\Delta_{\vk}^2)}.
\end{equation}
In the non-interacting case for $T=0$, we have therefore
\begin{equation}
  \underline {G}^0_d(\omega)^{-1}=\omega\unitop_{2}-\epsilon_{d}\tau_3
- \sum_{\vk}V_{\vk}^2\tau_3\underline {g}_{\vk}(i\omega_n)\tau_3.
\label{g0impsc}
\end{equation}
The local full Green's function matrix $\underline {G}_d(\omega)^{-1}$ for the
effective impurity model is given by the Dyson matrix equation 
\begin{equation}
\underline {G}_d(\omega)^{-1}= \underline  G_0^{-1}(\omega)- \underline
\Sigma(\omega),
\label{scgreenfct}
\end{equation}
where $\underline \Sigma(\omega)$ is the self-energy matrix.

\subsection{Self-energy using the higher $F$-Green's function}
As described by Bulla et al. \cite{BHP98} there is a method to calculate the
self-energy employing a higher $F$-Green's function, and it can also be used for the
case with superconducting bath. 
The calculation taking into account all offdiagonal terms yields
the following matrix equation 
\begin{equation}
\underline  G^0_d(\omega)^{-1} \underline {G}_d(\omega)-U\underline F(\omega)=\unitop_2,
\label{eomoff}
\end{equation}
with the matrix of higher Green's functions $\underline F(\omega)$,
\begin{equation}
  \underline F(\omega)=
\left(\begin{array}{c c}
F_{11}(\omega) & F_{12}(\omega) \\
F_{21}(\omega) & F_{22}(\omega)
\end{array}\right).
\end{equation}
We have introduced the matrix elements
$F_{11}(\omega)=\gfbraket{d_{\uparrow}n_{\downarrow};d^{\dagger}_{\uparrow}}_{\omega}$,
$F_{12}(\omega)=\gfbraket{d_{\uparrow}n_{\downarrow};d_{\downarrow}}_{\omega}$, 
$F_{21}(\omega)=-\gfbraket{d^{\dagger}_{\downarrow}n_{\uparrow};d^{\dagger}_{\uparrow}}_{\omega}$ and
$F_{22}(\omega)=-\gfbraket{d^{\dagger}_{\downarrow}n_{\uparrow};d_{\downarrow}}_{\omega}$.
In the NRG we calculate $F_{11}$ and $F_{21}$ and the others follow from
$F_{12}(\omega)=-F_{21}(-\omega)^*$ and $F_{22}(\omega)=F_{11}(-\omega)^*$.
We can define the self-energy matrix by
\begin{equation}
  \underline \Sigma(\omega)= U \underline F(\omega)\underline
  {G}_d(\omega)^{-1}.
\label{SigF}
\end{equation}
The properties of the Green's function and the higher $F$-Green's function
lead to the relations 
$\Sigma_{12}(\omega)=\Sigma_{21}(-\omega)^*$ and
$\Sigma_{22}(\omega)=-\Sigma_{11}(-\omega)^*$ for the self-energies.
We can therefore calculate the diagonal self-energy 
$\Sigma(\omega)=\Sigma_{11}(\omega)$ and the offdiagonal self-energy
$\Sigma^{\rm off}(\omega)=\Sigma_{21}(\omega)$ and deduce the other two matrix
elements from them.
With the relation (\ref{SigF}) between $\underline G$, $\underline F$ and
$\underline \Sigma$ the Dyson equation (\ref{scgreenfct}) is recovered from (\ref{eomoff}).
Therefore, once $G$ and $F$ are determined from the Lehmann representation the
self-energy can be calculated from (\ref{SigF}) and used in equations
(\ref{scselfcon}), (\ref{scgfct}) and (\ref{scgfctff}).

\section{Mapping of AFM and SC effective impurity model}
In the DMFT calculations with antiferromagnetic ordering the effective
impurity model can be given in the following discrete form
\begin{eqnarray}
H_{\rm AFM}=
\sum_{n,\alpha,{\sigma}}\xi_{n,\sigma}^{\alpha}a_{\alpha,n,\sigma}^{\dagger} a_{\alpha,n,\sigma}   
 +\sum_{n,\alpha,{\sigma}}\gamma^{\alpha}_{n,\sigma}(a_{\alpha,n,\sigma}^{\dagger}{d}_{\sigma}
 + \hc)
\nonumber
\label{afmAIMn}
\end{eqnarray}
where we have omitted the impurity term. Notice that the parameters are
$\sigma$-dependent. In this model the sublattice magnetic order is taken to
be in the $z$-direction,
whereas in the model with superconducting symmetry breaking (\ref{hamxigamsc2disc}) it corresponds to
a transverse direction, $x$ or $y$. Therefore we first perform a rotation in spin space
\begin{equation}
  a_{\alpha,n,\uparrow}\to\frac1{\sqrt2}(a_{\alpha,n,\uparrow}-a_{\alpha,n,\downarrow}),
\;\;
  a_{\alpha,n,\downarrow}\to\frac1{\sqrt2}(a_{\alpha,n,\uparrow}+a_{\alpha,n,\downarrow})
\end{equation}
and also for the $d$-operators.
This yields
\begin{eqnarray}
H_{\rm AFM}&=& \sum_{n,\alpha,{\sigma}}L_{n}^{\alpha}
a_{\alpha,n,\sigma}^{\dagger} a_{\alpha,n,\sigma}  +\sum_{n,\alpha,{\sigma}}
V_n^{\alpha}(a_{\alpha,n,\sigma}^{\dagger}{d}_{\sigma}+\hc)
\nonumber \\
&&-\sum_{n,\alpha}F_{n}^{\alpha}(a_{\alpha,n,\uparrow}^{\dagger}
a_{\alpha,n,\downarrow}+a_{\alpha,n,\downarrow}^{\dagger}
a_{\alpha,n,\uparrow})
 \nonumber\\ 
&& 
-\sum_{n,\alpha}  W_n^{\alpha}
(a_{\alpha,n,\uparrow}^{\dagger}d_{\downarrow}+a_{\alpha,n,\downarrow}^{\dagger}{d}_{\uparrow} 
+ \hc) \nonumber 
\end{eqnarray}
with
\begin{equation}
  L_{n}^{\alpha}=\frac{\xi_{n,\uparrow}^{\alpha}+\xi_{n,\downarrow}^{\alpha}}2,\;\;
V_n^{\alpha}=\frac{\gamma_{n,\uparrow}^{\alpha}+\gamma_{n,\downarrow}^{\alpha}}2,
\end{equation}
\begin{equation}
  F_{n}^{\alpha}=\frac{\xi_{n,\uparrow}^{\alpha}-\xi_{n,\downarrow}^{\alpha}}2,\;\;
  W_n^{\alpha}=\frac{\gamma_{n,\uparrow}^{\alpha}-\gamma_{n,\downarrow}^{\alpha}}2.
\nonumber
\end{equation}
Then we do a particle hole transformation for the down spin similar to (\ref{trafospinisospin}),
\begin{equation}
  a_{\alpha,n,\downarrow}\to a_{-\alpha,n,\downarrow}^{\dagger},
\qquad
d_{\downarrow}\to -d^{\dagger}_{\downarrow}.
\end{equation}
This gives
\begin{eqnarray}
H_{\rm AFM}&=& 
\sum_{n,\alpha}L_{n}^{\alpha}(a_{\alpha,n,\uparrow}^{\dagger} a_{\alpha,n,\uparrow}
+a_{-\alpha,n,\uparrow} a^{\dagger}_{-\alpha,n,\downarrow}) \nonumber\\
&&
+\sum_{n,\alpha}V_n^{\alpha}
(a^{\dagger}_{\alpha,n,\uparrow}{d}_{\uparrow}-a_{-\alpha,n,\downarrow}{d}^{\dagger}_{\downarrow}+\hc)
\nonumber \\
&&-\sum_{n,\alpha}
F_{n}^{\alpha}(a_{\alpha,n,\uparrow}^{\dagger}a^{\dagger}_{-\alpha,n,\downarrow}+a_{-\alpha,n,\downarrow} 
a_{\alpha,n,\uparrow})
 \nonumber\\ 
&& 
-\sum_{n,\alpha}
W_n^{\alpha}(-a_{\alpha,n,\uparrow}^{\dagger}d^{\dagger}_{\downarrow}+a_{-\alpha,n,\downarrow}{d}_{\uparrow} 
+ \hc) \nonumber 
\end{eqnarray}
So far we have made no assumption about the parameters
$\xi_{n,\sigma}^{\alpha}$, and $\gamma_{n,\sigma}^{\alpha}$. In the usual
scheme one has $\xi_{n,\sigma}^{-\alpha}=-\xi_{n,\sigma}^{\alpha}$, such that
$L_{n}^{-\alpha}=-L_{n}^{\alpha}$. Hence the second term in 
the first line is identical to the standard form apart from an additional
constant, when we use the fermionic anticommutation rules. In addition
$\xi_{n,\uparrow}^{\alpha}=\xi_{n,\downarrow}^{\alpha}$ is normally satisfied, such
that $F_{n}^{\alpha}=0$. Therefore the term in the third line, which looks
like the one for superconducting symmetry breaking, vanishes. We focus on the
half filling case where one additionally has  $\gamma_{n,\uparrow}^{\alpha}=\gamma_{n,\downarrow}^{-\alpha}$ 
So the other terms remain and one has a normal and an anomalous hopping term, 
\begin{eqnarray}
H_{\rm AFM}&=& \sum_{n,\alpha,\sigma}L_{n}^{\alpha}a_{\alpha,n,\sigma}^{\dagger} a_{\alpha,n,\sigma}
+\sum_{n,\alpha,\sigma}V_n^{\alpha}(a^{\dagger}_{\alpha,n,\sigma}{d}_{\sigma}+\hc)
 \nonumber\\
&&
+\sum_{n,\alpha}W_n^{\alpha}(a_{\alpha,n,\uparrow}^{\dagger}d^{\dagger}_{\downarrow}+{d}_{\uparrow}a_{\alpha,n,\downarrow}
+ \hc) \nonumber 
\end{eqnarray}
One can then do a Bogoliubov transformation,
\begin{equation}
\left(\begin{array}{c}
\! a_{\alpha,n,\uparrow} \\
\! a^{\dagger}_{\alpha,n,\downarrow}
\end{array}
\right )
 =
\left(\begin{array}{r r}
\! u_{n,\alpha} & \! -v_{n,\alpha}\\
\! v_{n,\alpha} & \! u_{n,\alpha}
\end{array}\right)
\left(\begin{array}{c}
\! b_{\alpha,n,\uparrow} \\
\! b^{\dagger}_{\alpha,n,\downarrow}
\end{array}\right),
\label{bogtrafo1}
\end{equation}
to obtain the desired Hamiltonian $\Hand^{\rm sc}$ in equation (\ref{hamxigamsc2disc}). 
The matrix elements are determined by
\begin{equation}
  u_{n,\alpha}^2-v_{n,\alpha}^2=\frac{V_n^{\alpha}\,{}^2-W_n^{\alpha}\,{}^2}{V_n^{\alpha}\,{}^2+W_n^{\alpha}\,{}^2},\;\;
  u_{n,\alpha}v_{n,\alpha}=\frac{-V_n^{\alpha}W_n^{\alpha}}{V_n^{\alpha}\,{}^2+W_n^{\alpha}\,{}^2}.
\end{equation}
The parameters $\xi_n^{\alpha},\gamma_n^{\alpha},\delta_n^{\alpha}$ in
(\ref{hamxigamsc2disc}) are related to the ones in $H_{\rm AFM}$ by 
\begin{equation}
  \xi_n^{\alpha}=(u_{n,\alpha}^2-v_{n,\alpha}^2)L_{n}^{\alpha}, \qquad
\gamma_n^{\alpha}=\sqrt{V_n^{\alpha}\,{}^2+W_n^{\alpha}\,{}^2},
\end{equation}
\begin{equation}
\delta_n^{\alpha}=-u_{n,\alpha}v_{n,\alpha}L_{n}^{\alpha}.
\end{equation}
We compared the numerical values obtained from the procedure described in
section III for the SC case with the ones from earlier AFM calculations for half
filling using the above relations. A reasonable agreement for the two
different calculations was found. 

\end{appendix}

\bibliography{artikel,biblio1,footnote}

\end{document}